\documentclass{article}
\usepackage{amssymb}

\usepackage{amsfonts}
\usepackage{amsmath}
\usepackage{graphicx}
\usepackage{amsmath, mathrsfs}
\usepackage{amssymb}
\usepackage{epsf}
\usepackage{psfrag}

\begin{document}

\newcounter{prob}\renewcommand{\theprob}{\arabic{prob}}

\title{Inflation: Homogeneous Limit}
\author{V. Mukhanov \\
ASC, Department of Physics, LMU, Munich}
\maketitle

\begin{abstract}
I review the motivation for the early-time cosmic acceleration stage in
expanding universe and discuss simple inflationary scenarios. Preheating and
reheating are considered in great detail. This is a sample chapter from my
book "Physical Foundations of Cosmology" published by Cambridge University
Press (2005).
\end{abstract}

Matter is distributed very homogeneously and isotropically on scales larger
than a few hundred Mpc. The microwave background gives us a
\textquotedblleft photograph\textquotedblright\ of the early universe, which
shows that at recombination the universe was extremely\textit{\ }homogeneous
and isotropic (with accuracy $\sim 10^{-4}$) on all scales up to the present
horizon. Given that the universe evolves according to the Hubble law, it is
natural to ask which initial conditions could lead to such homogeneity and
isotropy.

To obtain an exhaustive answer to this question we have to know the exact
physical laws which govern the evolution of the very early universe.
However, as long as we are interested only in the general features of the
initial conditions it suffices to know a few simple properties of these
laws. We will assume that inhomogeneity cannot be dissolved by expansion.
This natural surmise is supported by General Relativity (see Part II of this
book for details). We will also assume that nonperturbative quantum gravity
does not play an essential role at sub-Planckian curvatures. On the other
hand, we are nearly certain that nonperturbative quantum gravity effects
become very important when the curvature reaches Planckian values and the
notion of classical spacetime breaks down. Therefore we address the initial
conditions at the Planckian time $t_{i}=t_{Pl}\sim 10^{-43}$ s$.$

In this chapter we discuss the initial conditions problem we face in a
decelerating universe and show how this problem can be solved if the
universe undergoes a stage of the accelerated expansion known as inflation.

\section{Problem of initial conditions}

There are two \textit{independent} sets of initial conditions characterizing
matter: a) its spatial distribution, described by the energy density $%
\varepsilon \left( x\right) $ and b) the initial field of velocities. Let us
determine them given the current state of the universe.

\textbf{Homogeneity, isotropy (horizon) problem.}\textit{\ }The present
homogeneous, isotropic domain of the universe is at least as large as the
present horizon scale, $ct_{0}\sim 10^{28}$ cm$.$ Initially the size of this
domain was smaller by the ratio of the corresponding scale factors, $%
a_{i}/a_{0}$. Assuming that inhomogeneity cannot be dissolved by expansion,
we may safely conclude that the size of the homogeneous, isotropic region
from which our universe originated at $t=t_{i}$ was larger than
\begin{equation}
l_{i}\sim ct_{0}\frac{a_{i}}{a_{0}}.  \label{i1}
\end{equation}%
It is natural to compare this scale to the size of a causal region $%
l_{c}\sim ct_{i}:$%
\begin{equation}
\frac{l_{i}}{l_{c}}\sim \frac{t_{0}}{t_{i}}\frac{a_{i}}{a_{0}}.  \label{i2}
\end{equation}%
To obtain a rough estimate of this ratio we note that if the primordial
radiation dominates at $t_{i}\sim t_{Pl},$ then its temperature is $%
T_{Pl}\sim 10^{32}$ K$.$ Hence%
\begin{equation*}
\left( a_{i}/a_{0}\right) \sim \left( T_{0}/T_{Pl}\right) \sim 10^{-32}
\end{equation*}%
and we obtain
\begin{equation}
\frac{l_{i}}{l_{c}}\sim \frac{10^{17}}{10^{-43}}10^{-32}\sim 10^{28}.
\label{i3}
\end{equation}%
Thus, at the initial Planckian time, the size of our universe exceeded the
causality scale by $28$ orders of magnitude. This means that in $10^{84}$
\textit{causally disconnected} regions the energy density was smoothly
distributed\ with fractional variation not exceeding $\delta \varepsilon
/\varepsilon \sim 10^{-4}.$ Because no signals can propagate faster than
light, no causal physical processes can be responsible for such an
unnaturally fine-tuned matter distribution.

Assuming that the scale factor grows as some power of time, we can use an
estimate $a/t\sim \dot{a}$ and rewrite formula (\ref{i2}) as
\begin{equation}
\frac{l_{i}}{l_{c}}\sim \frac{\dot{a}_{i}}{\dot{a}_{0}}.  \label{i4}
\end{equation}%
Thus, the size of our universe was initially larger than that of a causal
patch by the ratio of the corresponding expansion rates. Assuming that
gravity was always attractive and hence was decelerating the expansion, we
conclude from (\ref{i4}) that the homogeneity scale was always larger than
the scale of causality. Therefore, the homogeneity problem is also sometimes
called the \textit{horizon} problem.

\textbf{Initial velocities (flatness) problem.}\textit{\ }Let us suppose for
a minute that someone has managed to distribute matter in the required way.
The next question concerns initial velocities. Only after they are specified
is the Cauchy problem completely posed and can the equations of motion be
used to predict the future of the universe unambiguously. The initial
velocities must obey the Hubble law because otherwise the initial
homogeneity is very quickly spoiled. That this has to occur in so many
causally disconnected regions further complicates the horizon problem.
Assuming that it has, nevertheless, been achieved,\ we can ask how
accurately the initial Hubble velocities have to be chosen for a given
matter distribution.

Let us consider a large spherically symmetric cloud of matter and compare
its total energy with the kinetic energy due to Hubble expansion, $E^{k}.$
The total energy is the sum of the positive kinetic energy and the negative
potential energy of the gravitational self-interaction, $E^{p}.$ It is
conserved:%
\begin{equation*}
E^{tot}=E_{i}^{k}+E_{i}^{p}=E_{0}^{k}+E_{0}^{p}.
\end{equation*}%
Because the kinetic energy is proportional to the velocity squared,
\begin{equation*}
E_{i}^{k}=E_{0}^{k}\left( \dot{a}_{i}/\dot{a}_{0}\right) ^{2}
\end{equation*}%
and we have
\begin{equation}
\frac{E_{i}^{tot}}{E_{i}^{k}}=\frac{E_{i}^{k}+E_{i}^{p}}{E_{i}^{k}}=\frac{%
E_{0}^{k}+E_{0}^{p}}{E_{0}^{k}}\left( \frac{\dot{a}_{0}}{\dot{a}_{i}}\right)
^{2}.  \label{i5}
\end{equation}%
Since $E_{0}^{k}\sim \left\vert E_{0}^{p}\right\vert $ and $\dot{a}_{0}/\dot{%
a}_{i}\leq 10^{-28},$ we find
\begin{equation}
\frac{E_{i}^{tot}}{E_{i}^{k}}\leq 10^{-56}.  \label{i6}
\end{equation}%
This means that for a given energy density distribution the initial Hubble
velocities must be adjusted so that the huge negative gravitational energy
of the matter is compensated by a huge positive kinetic energy to an
unprecedented accuracy of $10^{-54}\%.$ An error\ in the initial velocities
exceeding $10^{-54}\%$ has a dramatic consequence: the universe either
recollapses or becomes \textquotedblleft empty\textquotedblright\ too early.
To stress the unnaturalness of this requirement one speaks of the initial
velocities problem.

\begin{description}
\item \addtocounter{prob}{+1} \textbf{Problem \theprob. }How can the above
consideration be made rigorous using the Birkhoff theorem?
\end{description}

In general relativity the problem described can be reformulated in terms of
the cosmological parameter $\Omega \left( t\right) $. Using the definition
of $\Omega \left( t\right) $ we can rewrite Friedmann equation as
\begin{equation}
\Omega \left( t\right) -1=\frac{k}{\left( Ha\right) ^{2}},  \label{i7}
\end{equation}%
and hence
\begin{equation}
\Omega _{i}-1=\left( \Omega _{0}-1\right) \frac{\left( Ha\right) _{0}^{2}}{%
\left( Ha\right) _{i}^{2}}=\left( \Omega _{0}-1\right) \left( \frac{\dot{a}%
_{0}}{\dot{a}_{i}}\right) ^{2}\leq 10^{-56}.  \label{i8}
\end{equation}%
Note that this relation immediately follows from (\ref{i5}) if we take into
account that $\Omega =\left\vert E^{p}\right\vert /E^{k}.$ We infer from (%
\ref{i8}) that the cosmological parameter must initially be extremely close
to unity$,$ corresponding to a \textit{flat} universe. Therefore the problem
of initial velocities is also called the \textit{flatness }problem.

\textbf{Initial perturbation problem.} One further problem we mention here
for completeness is the origin of the primordial inhomogeneities needed to
explain the large-scale structure of the universe. They must be initially of
order $\delta \varepsilon /\varepsilon \sim 10^{-5}$ on galactic scales.
This further aggravates the very difficult problem of homogeneity and
isotropy, making it completely intractable. We will see later that the
problem of initial perturbations has the same roots as the horizon and
flatness problems and that it can also be successfully solved in
inflationary cosmology. However, for the moment we put it aside and proceed
with the ``more easy''\ problems.

The above considerations clearly show that the initial conditions which led
to the observed universe are very unnatural and non-generic. Of course, one
can make the objection that naturalness is a question\ of taste and even
claim that the most simple and symmetric initial conditions are ``more
physical.''\ In the absence of a\ quantitative measure of ``naturalness''
for a set of initial conditions it is very difficult to argue with this
attitude. On the other hand it is hard to imagine any measure which selects
the special and degenerate conditions in preference to the generic ones. In
the particular case under consideration the generic conditions would mean
that the initial distribution of the matter is strongly inhomogeneous with $%
\delta \varepsilon /\varepsilon \gtrsim 1$ everywhere or, at least, in the
causally disconnected regions.

The universe is unique and we do not have the opportunity to repeat the
\textquotedblleft experiment of creation\textquotedblright\ many times.
Therefore cosmological theory can claim to be a successful physical theory
only if it can explain the state of the observed universe using simple
physical ideas and starting with the most generic initial conditions.
Otherwise it would simply amount to \textquotedblleft cosmic
archaeology,\textquotedblright\ where \textquotedblleft cosmic
history\textquotedblright\ is written on the basis of a limited number of
hot big bang remnants.\ If we are pretentious enough to answer the question
raised by Einstein, \textquotedblleft What really interests me is whether
God had any choice when he created the World,\textquotedblright\ we must be
able to explain how a particular universe can be created starting with
generic initial conditions. The inflationary paradigm seems to be a step in
the right direction and it strongly restricts \textquotedblleft God's
choice.\textquotedblright\ Moreover, it makes important predictions which
can be verified experimentally (observationally), thus giving cosmology the
status of a physical theory.

\section{Inflation: main idea}

We have seen so far that the same ratio, $\dot{a}_{i}/\dot{a}_{0}$, enters
both sets of \textit{independent} initial conditions$.$ The large value of
this ratio determines the number of causally disconnected regions and
defines the necessary accuracy of the initial velocities. If gravity was
always attractive, then $\dot{a}_{i}/\dot{a}_{0}$ is necessarily larger than
unity because gravity decelerates an expansion. Therefore, the conclusion $%
\dot{a}_{i}/\dot{a}_{0}\gg 1$ can be avoided \textit{only} if we assume that
during some period of expansion gravity acted as a \textquotedblleft
repulsive\textquotedblright\ force, thus accelerating the expansion. \ In
this case we can have $\dot{a}_{i}/\dot{a}_{0}<1$ and the creation of our
type of universe from a single causally connected domain may become
possible. A period of accelerated expansion is a \textit{necessary}
condition, but whether is it also sufficient depends on the particular model
where this condition is realized. With these remarks in mind we arrive at
the following general definition of inflation:

\textit{Inflation is a stage of accelerated expansion of the universe when
gravity acts as a repulsive force.}

\begin{figure}
\begin{center}
\psfrag{y}[tr]{\small$\dot{a}$}
\psfrag{x}[tr]{\small$t$}
\psfrag{q}[]{$\mathbf{?}$}
\psfrag{graceful exit}[t]{\small graceful exit}
\psfrag{Inflation}[lB]{\small Inflation}
\psfrag{decelerated Friedmann expansion}[]{\small decelerated Friedmann expansion}
\includegraphics[width=3in,angle=0]{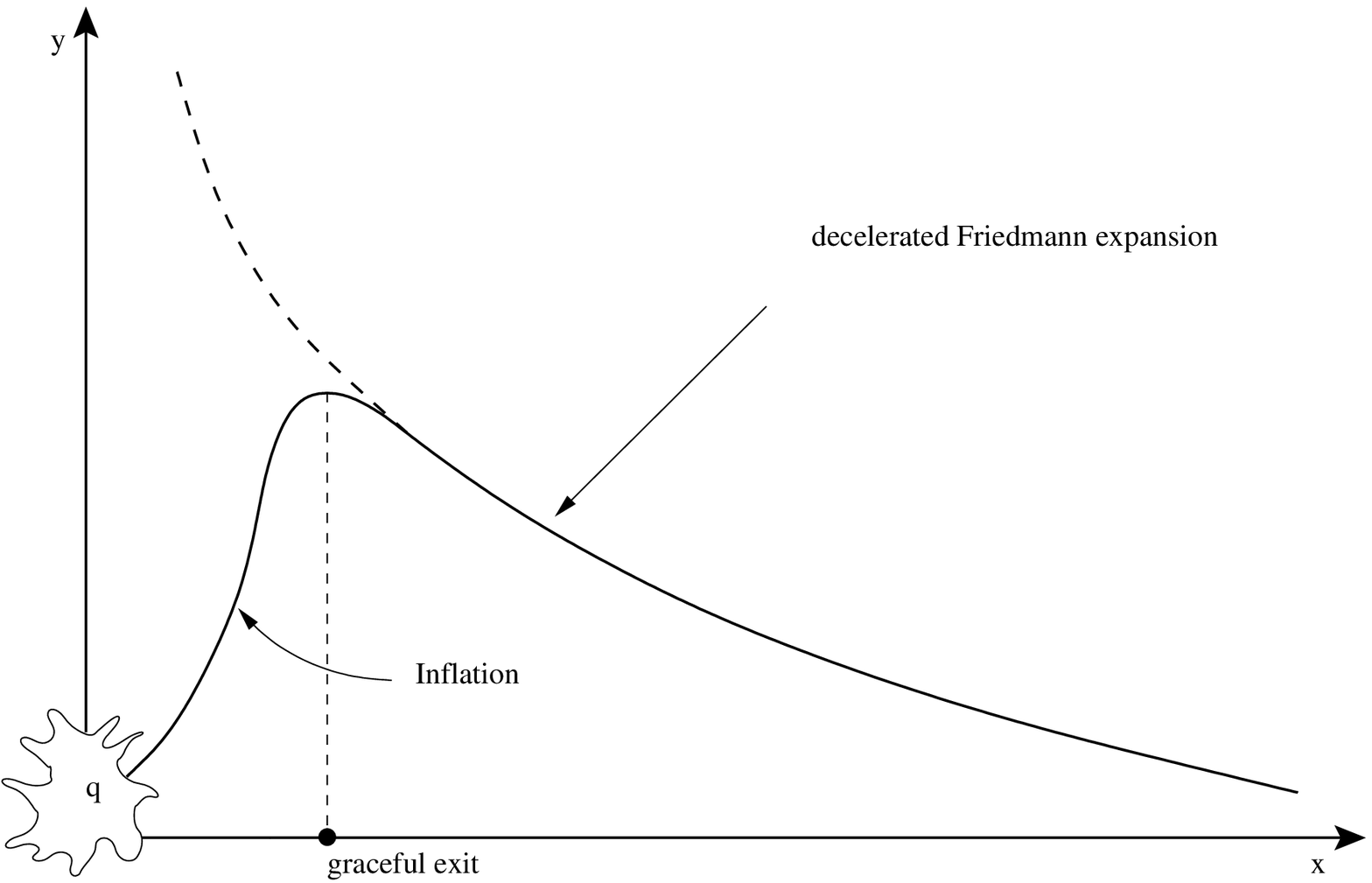}
\caption{~}
\end{center}
\end{figure}

Figure 1 shows how the old\ picture of a decelerated Friedmann universe is
modified by inserting a stage of cosmic acceleration. It is obvious that if
we do not want to spoil the successful predictions of the standard Friedmann
model, such as nucleosynthesis, inflation should begin and end sufficiently
early. We will see later that the requirement of the generation of
primordial fluctuations further restricts the energy scale of inflation;
namely, in the simple models inflation should be over at $t_{f}$ $\sim
10^{-34}$ to $10^{-36}$ s$.$ Successful inflation must also possess a smooth
graceful exit into the decelerated Friedmann stage because otherwise the
homogeneity of the universe would be destroyed.

Inflation explains the origin of the Big Bang; since it accelerates the
expansion, small initial velocities within a causally connected patch become
very large. Furthermore, inflation can produce the whole observable universe
from a small homogeneous domain even if the universe was strongly
inhomogeneous outside of this domain. The reason is that in an accelerating
universe there \textit{always} exists an event horizon. It has size
\begin{equation}
r_{e}\left( t\right) =a\left( t\right) \int\limits_{t}^{t_{\max }}\frac{dt}{a%
}=a\left( t\right) \int\limits_{a\left( t\right) }^{a_{\max }}\frac{da}{\dot{%
a}a}.  \label{ievhor}
\end{equation}%
The integral converges even if $a_{\max }\rightarrow \infty $ because the
expansion rate $\dot{a}$ grows with $a.$ The existence of an event horizon
means that anything at time $t$ a distance larger than $r_{e}\left( t\right)
$ from an observer cannot influence that observer's future. Hence the future
evolution of the region inside a ball of radius $r_{e}\left( t\right) $ is
completely independent of the conditions outside a ball of radius $%
2r_{e}\left( t\right) $ centered at the same place. Let us assume that at $%
t=t_{i}$ matter was distributed homogeneously and isotropically only inside
a ball of radius $2r_{e}\left( t_{i}\right) $ (Fig.~2). Then an
inhomogeneity propagating from outside this ball can spoil the homogeneity
only in the region which was initially between the spheres of radii $%
r_{e}\left( t_{i}\right) $ and $2r_{e}\left( t_{i}\right) .$ The region
originating from the sphere of radius $r_{e}\left( t_{i}\right) $ remains
homogeneous. This internal domain can be influenced only by events which
happened at $t_{i}$ between the two spheres, where the matter was initially
distributed homogeneously and isotropically.

\begin{figure}
\begin{center}
\psfrag{R}[]{\small$2r_{e}(t_{i})$}
\psfrag{A}[]{\small$r_{e}(t_{i}){\displaystyle\frac{a_{f}}{a_{i}}}$}
\psfrag{H}[b]{\small Homogeneity}
\psfrag{I}[]{\small is}
\psfrag{P}[t]{\small preserved}
\includegraphics[width=0.9\textwidth,angle=0]{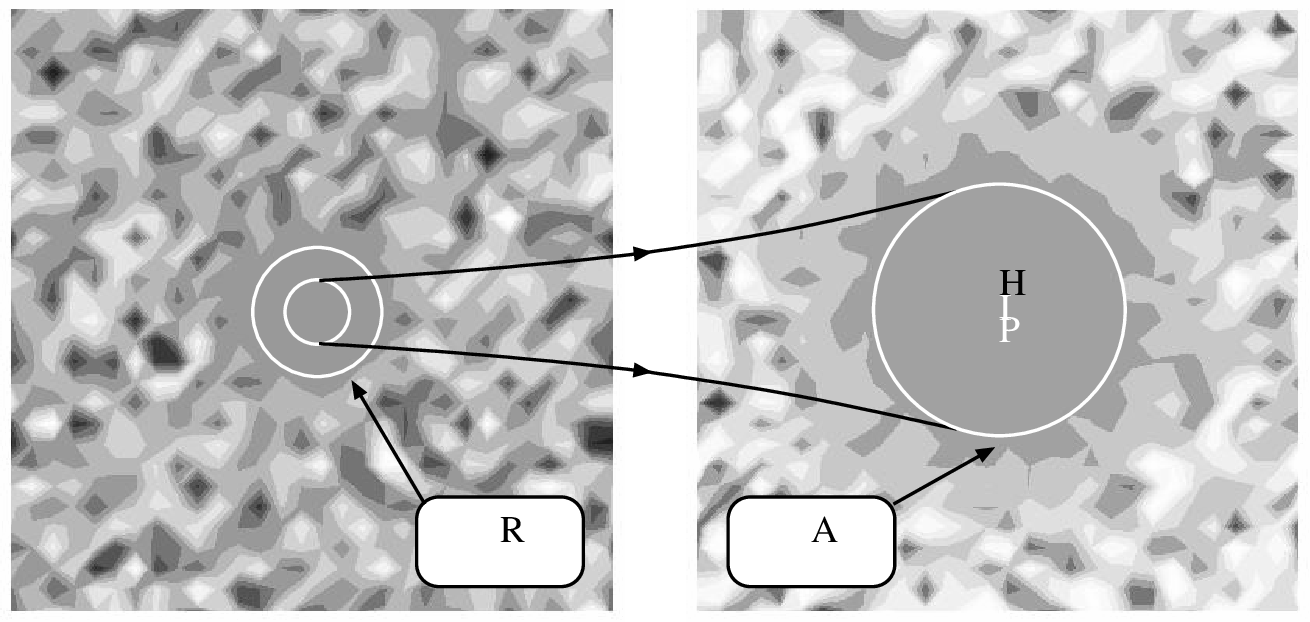}
\caption{~}
\end{center}
\end{figure}

The physical size of the homogeneous internal region increases and is equal
to
\begin{equation}
r_{h}\left( t_{f}\right) =r_{e}\left( t_{i}\right) \frac{a_{f}}{a_{i}}
\label{ih}
\end{equation}%
at the end of inflation. It is natural to compare this scale with the
particle horizon size, which in an accelerated universe can be estimated as
\begin{equation}
r_{p}\left( t\right) =a\left( t\right) \int\limits_{t_{i}}^{t}\frac{dt}{a}%
=a\left( t\right) \int\limits_{a_{i}}^{a}\frac{da}{\dot{a}a}\sim \frac{%
a\left( t\right) }{a_{i}}r_{e}\left( t_{i}\right) ,  \label{ip}
\end{equation}%
since the main contribution to the integral comes from $a\sim a_{i}.$ At the
end of inflation $r_{p}\left( t_{f}\right) \sim r_{h}\left( t_{f}\right) ,$
that is, the size of the homogeneous region, originating from a causal
domain, is of order the particle horizon scale.

Thus, instead of considering a homogeneous universe in many causally
disconnected regions, we can begin with a small homogeneous causal domain
which inflation blows up to a very large size, preserving the homogeneity
irrespective of the conditions outside this domain.

\begin{description}
\item \addtocounter{prob}{+1} \textbf{Problem \theprob. }Why does the above
consideration fail in a decelerating universe?
\end{description}

The next question is whether we can relax the restriction of homogeneity on
the initial conditions. Namely, if we begin with a strongly inhomogeneous
causal domain, can inflation still produce a large homogeneous universe?

The answer to this question is positive. Let us assume that the initial
energy density inhomogeneity is of order unity on scales $\sim H_{i}^{-1},$
that is,
\begin{equation}
\left( \frac{\delta \varepsilon }{\varepsilon }\right) _{t_{i}}\sim \frac{1}{%
\varepsilon }\frac{\left\vert \nabla \varepsilon \right\vert }{a_{i}}%
H_{i}^{-1}=\frac{\left\vert \nabla \varepsilon \right\vert }{\varepsilon }%
\frac{1}{\dot{a}_{i}}\sim O\left( 1\right) ,  \label{ibinh}
\end{equation}%
where $\nabla $ is the spatial derivative with respect to the comoving
coordinates. At $t\gg t_{i},$ the contribution of this inhomogeneity to the
variation of the energy density within the Hubble scale $H\left( t\right)
^{-1}$ can be estimated as
\begin{equation}
\left( \frac{\delta \varepsilon }{\varepsilon }\right) _{t}\sim \frac{1}{%
\varepsilon }\frac{\left\vert \nabla \varepsilon \right\vert }{a\left(
t\right) }H\left( t\right) ^{-1}\sim O\left( 1\right) \frac{\dot{a}_{i}}{%
\dot{a}\left( t\right) },  \label{ilinh}
\end{equation}%
where we have assumed that $\left\vert \nabla \varepsilon \right\vert
/\varepsilon $ does not change substantially during expansion. This
assumption is supported by the analysis of the behavior of linear
perturbations on scales larger than the curvature scale $H^{-1}$ (see
chapters 7 and 8). It follows from (\ref{ilinh}) that if the universe
undergoes a stage of acceleration, that is, $\dot{a}\left( t\right) >\dot{a}%
_{i}$ for $t>t_{i},$ then the contribution of a large initial inhomogeneity
to the energy variation on the curvature scale disappears. A patch of size $%
H^{-1}$ becomes more and more homogeneous because the initial inhomogeneity
is \textquotedblleft kicked out\textquotedblright : the physical size of the
perturbation, $\propto a$, grows faster than the curvature scale, $H^{-1}=a/%
\dot{a}$, while the perturbation amplitude does not change substantially.
Since inhomogeneities are \textquotedblleft devalued\textquotedblright\
within the curvature scale,\ the name \textquotedblleft
inflation\textquotedblright\ fairly captures the physical effect of
accelerated expansion. The consideration above is far from rigorous.
However, it gives the flavor\ of the \textquotedblleft
no-hair\textquotedblright\ theorem for an inflationary stage.

To sum up, inflation demolishes large initial inhomogeneities and produces a
homogeneous, isotropic domain. It follows from (\ref{ilinh}) that if we want
to avoid the situation of a large initial perturbation reentering the
present horizon, $\sim H_{0}^{-1},$ and inducing a large inhomogeneity, we
have to assume that the initial expansion rate was \textit{much} \textit{%
smaller} than the rate of expansion today, that is, $\dot{a}_{i}/\dot{a}%
_{0}\ll 1.$ More precisely, the CMB observations require that the variation
of the energy density on the present horizon scale does not exceed $10^{-5}.$
The traces of an initial large inhomogeneity will be sufficiently diluted
only if $\dot{a}_{i}/\dot{a}_{0}<10^{-5}.$ Rewriting relation (\ref{i8}) as
\begin{equation}
\Omega _{0}=1+\left( \Omega _{i}-1\right) \left( \frac{\dot{a}_{i}}{\dot{a}%
_{0}}\right) ^{2},  \label{i10}
\end{equation}%
we see that if $\left\vert \Omega _{i}-1\right\vert \sim O\left( 1\right) $
then
\begin{equation}
\Omega _{0}=1
\end{equation}%
to very high accuracy. This important \textit{robust prediction} of
inflation has a kinematical origin and it states that the \textit{total }%
energy density of all components of matter, \textit{irrespective of their
origin}, must be equal to the critical energy density today. We will see
later that amplified quantum fluctuations lead to the tiny corrections to $%
\Omega _{0}=1$ , which are of order $10^{-5}.$ It is worth noting that, in
contrast to a decelerating universe where $\Omega \left( t\right)
\rightarrow 1$ as $t\rightarrow 0,$ in an accelerating universe $\Omega
\left( t\right) \rightarrow 1$ as $t\rightarrow \infty ,$ that is, $\Omega
=1 $ is its \textit{future} attractor.

\begin{description}
\item \addtocounter{prob}{+1} \textbf{Problem \theprob. }Why does the
consideration above fail for $\Omega _{i}=0$?
\end{description}

\section{How can gravity become ``repulsive''?}

To answer this question we recall the Friedmann equation:
\begin{equation}
\text{\ }\ddot{a}=-\frac{4\pi }{3}G(\varepsilon +3p)a.  \label{iacceq}
\end{equation}%
Obviously, if the strong energy dominance condition, $\varepsilon +3p>0,$ is
satisfied, then $\ddot{a}$ $<0$ and gravity decelerates the expansion. The
universe can undergo a stage of accelerated expansion with $\ddot{a}$ $>0$
only if this condition is violated, that is, if $\varepsilon +3p<0$. One
particular example of \textquotedblleft matter\textquotedblright\ with
broken energy dominance condition is a positive cosmological constant, for
which $p_{V}=-\varepsilon _{V}$ and $\varepsilon +3p=-2\varepsilon _{V}<0.$
In this case the solution of Einstein's equations is a de Sitter universe
discussed in previous Chapters. For $t\gg H_{\Lambda }^{-1},$ the de Sitter
universe expands exponentially quickly, $a\propto \exp \left( H_{\Lambda
}t\right) ,$ and the rate of expansion grows as the scale factor$.$ The
\textit{exact} de Sitter solution fails to satisfy all necessary conditions
for successful inflation: namely, it does not possess a smooth graceful exit
into the Friedmann stage. Therefore, in realistic inflationary models, it
can be utilized only as a zero order approximation. To have a graceful exit
from inflation we must allow the Hubble parameter to vary in time$.$

Let us now determine the general conditions which must be satisfied in a
successful inflationary model. Because
\begin{equation}
\frac{\ddot{a}}{a}=H^{2}+\dot{H},  \label{i12}
\end{equation}
and $\ddot{a}$ should become negative during a graceful exit, the derivative
of the Hubble constant, $\dot{H},$ must obviously be negative. The ratio $%
\left| \dot{H}\right| /H^{2}$ grows toward the end of inflation and the
graceful exit takes place when $\left| \dot{H}\right| $ becomes of order $%
H^{2}.$ Assuming that $H^{2}$ changes faster than $\dot{H}$ , that is, $%
\left| \ddot{H}\right| <2H\dot{H},$ we obtain the following generic estimate
for the duration of inflation:
\begin{equation}
t_{f}\sim H_{i}/\left| \dot{H}_{i}\right| ,  \label{idur}
\end{equation}
where $H_{i}$ and $\dot{H}_{i}$ refer to the beginning of inflation. At $%
t\sim t_{f}$ the expression on the right hand side in equation (\ref{i12})
changes sign and the universe begins to decelerate.

Inflation should last long enough to stretch a small domain to the scale of
the observable universe. Rewriting the condition $\dot{a}_{i}/\dot{a}%
_{0}<10^{-5}$ as
\begin{equation*}
\frac{\dot{a}_{i}}{\dot{a}_{f}}\frac{\dot{a}_{f}}{\dot{a}_{0}}=\frac{a_{i}}{%
a_{f}}\frac{H_{i}}{H_{f}}\frac{\dot{a}_{f}}{\dot{a}_{0}}<10^{-5},
\end{equation*}%
and taking into account that $\dot{a}_{f}/\dot{a}_{0}$ should be larger than
$10^{28},$ we conclude that inflation is successful only if
\begin{equation*}
\frac{a_{f}}{a_{i}}>10^{33}\frac{H_{i}}{H_{f}}.
\end{equation*}%
Let us assume that $\left\vert \dot{H}_{i}\right\vert \ll H_{i}^{2}$ and
neglect the change of the Hubble parameter. Then the ratio of the scale
factors can be roughly estimated as
\begin{equation}
a_{f}/a_{i}\sim \exp \left( H_{i}t_{f}\right) \sim \exp \left(
H_{i}^{2}/\left\vert \dot{H}_{i}\right\vert \right) >10^{33}.  \label{i14a}
\end{equation}%
Hence inflation can solve the initial conditions problem only if $t_{f}>75$ $%
H_{i}^{-1},$ that is, it lasts \textit{longer} than seventy five Hubble
times (e-folds). Rewritten in terms of the initial values of the Hubble
parameter and its derivative, this condition takes the form
\begin{equation}
\frac{\left\vert \dot{H}_{i}\right\vert }{H_{i}^{2}}<\frac{1}{75}.
\label{i15}
\end{equation}%
Using the Friedmann equations, we can reformulate it in terms of the bounds
on the initial equation of state
\begin{equation}
\frac{\left( \varepsilon +p\right) _{i}}{\varepsilon _{i}}<10^{-2}.
\label{i16}
\end{equation}%
Thus, at the beginning of inflation the deviation from the vacuum equation
of state must not exceed one percent. Therefore an exact de Sitter solution
is a very good approximation for the initial stage of inflation. Inflation
ends when $\varepsilon +p\sim \varepsilon $.

\begin{description}
\item \addtocounter{prob}{+1} \textbf{Problem \theprob.} Consider an
exceptional case where $\left\vert \dot{H}\right\vert $ decays at the same
rate as $H^{2},$ that is, $\dot{H}=-pH^{2},$ where $p=const.$ Show that for $%
p<1$ we have power-law inflation. This inflation has no natural graceful
exit and in this sense is similar to a pure de Sitter universe.
\end{description}

\section{How to realize the equation of state $p\approx -\protect\varepsilon
$}

Thus far we have used the language of ideal hydrodynamics, which is an
adequate phenomenological description of matter on large scales. Now we
discuss a simple field-theoretic model where the required equation of state
can be realized. The natural candidate to drive inflation is a scalar field.
The name given to such a field is the \textquotedblleft
inflaton.\textquotedblright\ We saw that the energy-momentum tensor for a
scalar field can be rewritten in a form which mimics an ideal fluid. The
homogeneous classical field (scalar condensate) is then characterized by
energy density
\begin{equation}
\varepsilon =\frac{1}{2}\dot{\varphi}^{2}+V\left( \varphi \right) ,
\label{ien}
\end{equation}%
and pressure
\begin{equation}
p=\frac{1}{2}\dot{\varphi}^{2}-V\left( \varphi \right) .  \label{ipr}
\end{equation}%
We have neglected spatial derivatives here because they become negligible
soon after the beginning of inflation due to the \textquotedblleft
no-hair\textquotedblright\ theorem.

\begin{description}
\item \addtocounter{prob}{+1} \textbf{Problem \theprob. }Consider a massive
scalar field with potential $V=\frac{1}{2}m^{2}\varphi ^{2},$ where $m\ll
m_{Pl},$ and determine the bound on the allowed inhomogeneity imposed by the
requirement that the energy density must not exceed the Planckian value. Why
does the contribution of the spatial gradients to the EMT decay more quickly
than the contribution of the mass term?
\end{description}

It follows from equations (\ref{ien}) and (\ref{ipr}) that the scalar field
has the desired equation of state only if $\dot{\varphi}^{2}\ll V\left(
\varphi \right) .$ Because $p=-\varepsilon +\dot{\varphi}^{2}$, the
deviation of the equation of state from that for the vacuum is entirely
characterized by the kinetic energy,\ $\dot{\varphi}^{2},$ which must be
much smaller than the potential energy $V\left( \varphi \right) .$
Successful realization of inflation thus requires keeping $\dot{\varphi}^{2}$
small compared to $V\left( \varphi \right) $ during a sufficiently long time
interval, or more precisely, for at least 75 e-folds. In turn this depends
on the shape of the potential $V\left( \varphi \right) $. To determine which
potentials can provide us with inflation, we have to study the behavior of a
homogeneous classical scalar field in an expanding universe. The equation
for this field can be derived either directly from the Klein-Gordon equation
or by substituting (\ref{ien}) and (\ref{ipr}) into the conservation law $%
\dot{\varepsilon}=-3H\left( \varepsilon +p\right) $. The result is
\begin{equation}
\ddot{\varphi}+3H\dot{\varphi}+V_{,\varphi }=0,  \label{iscfeq}
\end{equation}%
where $V_{,\varphi }\equiv \partial V/\partial \varphi .$ This equation has
to be supplemented by the Friedmann equation:
\begin{equation}
H^{2}=\frac{8\pi }{3}\left( \frac{1}{2}\dot{\varphi}^{2}+V\left( \varphi
\right) \right) ,  \label{ifreq}
\end{equation}%
where we have set $G=1$ and $k=0.$ We first find the solutions of equations (%
\ref{iscfeq}) and (\ref{ifreq}) for a free massive scalar field and then
study the behavior of the scalar field in the case of a general potential $%
V\left( \varphi \right) .$

\subsection{Simple example: $V=\dfrac{1}{2}m^{2}\protect\varphi ^{2}.$}

Substituting $H$ from (\ref{ifreq}) into (\ref{iscfeq}), we obtain the
closed form equation for $\varphi ,$%
\begin{equation}
\ddot{\varphi}+\sqrt{12\pi }\left( \dot{\varphi}^{2}+m^{2}\varphi
^{2}\right) ^{1/2}\dot{\varphi}+m^{2}\varphi =0.  \label{iscfn}
\end{equation}%
This is a nonlinear second order differential equation with no explicit time
dependence. Therefore it can be reduced to a first order differential
equation for $\dot{\varphi}\left( \varphi \right) .$ Taking into account
that
\begin{equation*}
\ddot{\varphi}=\dot{\varphi}\frac{d\dot{\varphi}}{d\varphi },
\end{equation*}%
equation (\ref{iscfn}) becomes
\begin{equation}
\frac{d\dot{\varphi}}{d\varphi }=-\frac{\sqrt{12\pi }\left( \dot{\varphi}%
^{2}+m^{2}\varphi ^{2}\right) ^{1/2}\dot{\varphi}+m^{2}\varphi }{\dot{\varphi%
}},  \label{ifoeq}
\end{equation}%
which can be studied using the phase diagram method. The behavior of the
solutions in $\varphi -\dot{\varphi}$ plane is shown in Figure 3. \ The
important feature of this diagram is the existence of an attractor solution
to which all other solutions converge in time. One can distinguish different
regions corresponding to different effective equations of state. Let us
consider them in more detail. We restrict ourselves to the lower right
quadrant ($\varphi >0,$ $\dot{\varphi}<0$); solutions in the other quadrants
can easily be derived simply by taking into account the symmetry of the
diagram.

\begin{figure}
\begin{center}
\psfrag{x}[br]{\small$\varphi$}
\psfrag{y}[tr]{\small$\dot{\varphi}$}
\psfrag{A}[bl]{\small Attractor}
\psfrag{m}[bl]{\small$\frac{m}{\sqrt{12\pi}}$}
\psfrag{-m}[tr]{\small$\frac{-m}{\sqrt{12\pi}}$}
\includegraphics[width=0.95\textwidth,angle=0]{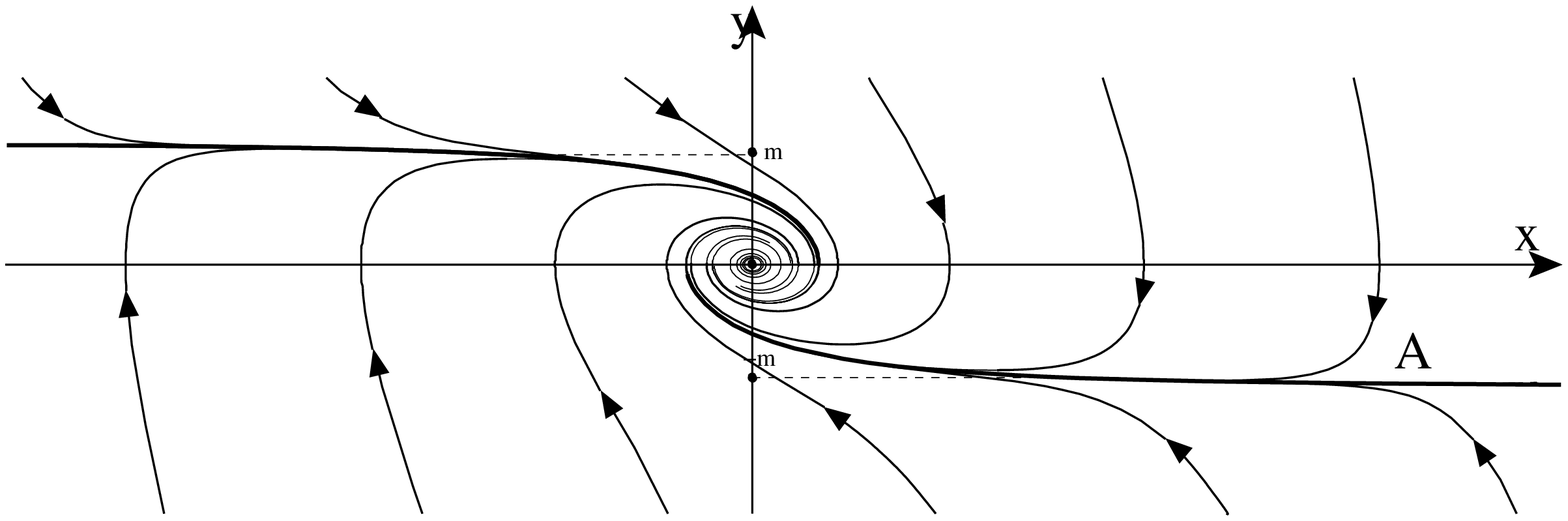}
\caption{~}
\end{center}
\end{figure}

\textbf{Ultra-hard equation of state.}\textit{\ }First we study the region
where $\left\vert \dot{\varphi}\right\vert \gg m\varphi .$ It describes the
situation when the potential energy is small compared to the kinetic energy,
so that $\dot{\varphi}^{2}\gg V.$ It follows from equations (\ref{ien}) and (%
\ref{ipr}) that in this case the equation of state is ultra-hard, $p\approx
+\varepsilon .$ Neglecting $m\varphi $ compared to $\dot{\varphi}$ in
equation (\ref{ifoeq}), we obtain
\begin{equation}
\frac{d\dot{\varphi}}{d\varphi }\simeq \sqrt{12\pi }\dot{\varphi}.
\label{ired1}
\end{equation}%
The solution of this equation is
\begin{equation}
\dot{\varphi}=C\exp \left( \sqrt{12\pi }\varphi \right) ,  \label{isol1}
\end{equation}%
where $C<0$ is a constant of integration. In turn, solving (\ref{isol1}) for
$\varphi \left( t\right) $ gives
\begin{equation}
\varphi =const-\frac{1}{\sqrt{12\pi }}\ln t.  \label{isol2}
\end{equation}%
Substituting this result into equation (\ref{ifreq}) and neglecting the
potential term, we obtain
\begin{equation}
H^{2}\equiv \left( \frac{\dot{a}}{a}\right) ^{2}\simeq \frac{1}{9t^{2}}.
\label{ieq}
\end{equation}%
It immediately follows that $a\propto t^{1/3}$ and $\varepsilon \propto
a^{-6}$ in agreement with the ultra-hard equation of state. Note that the
solution obtained is exact for a massless scalar field. According to (\ref%
{isol1}) the derivative of the scalar field decays exponentially more
quickly than the value of the scalar field itself. Therefore, the large
initial value of $\left\vert \dot{\varphi}\right\vert $ is damped within a
short time interval before the field $\varphi $ itself has changed
significantly. The trajectory which begins at large $\left\vert \dot{\varphi}%
\right\vert $ goes up very sharply and meets the attractor. This
substantially enlarges the set of initial conditions which lead to an
inflationary stage.

\textbf{Inflationary solution.} If a trajectory joins\ the attractor where
it is flat, at $\left\vert \varphi \right\vert \gg 1,$ then afterwards the
solution describes a stage of accelerated expansion (recall that we work in
Planckian units). To determine the attractor solution we assume that $d\dot{%
\varphi}/d\varphi \approx 0$ along its trajectory. It follows from (\ref%
{ifoeq}) $\ $that
\begin{equation}
\dot{\varphi}_{\text{atr}}\approx -\frac{m}{\sqrt{12\pi }},  \label{iatt1}
\end{equation}%
and therefore
\begin{equation}
\varphi _{\text{atr}}\left( t\right) \simeq \varphi _{i}-\frac{m}{\sqrt{%
12\pi }}(t-t_{i})\simeq \frac{m}{\sqrt{12\pi }}\left( t_{f}-t\right) ,
\label{isol3}
\end{equation}%
where $t_{i}$ is the time when the trajectory joins the attractor and $t_{f}$
is the moment when $\varphi $ \textit{formally} vanishes. In reality,
solution (\ref{isol3}) fails well before the field $\varphi $ vanishes.

\begin{description}
\item \addtocounter{prob}{+1} \textbf{Problem \theprob. }Calculate the
corrections to the approximate attractor solution (\ref{iatt1}) and show
that
\index{inflation!attractor solution}
\begin{equation}
\dot{\varphi}_{\text{atr}}=-\frac{m}{\sqrt{12\pi }}\left( 1-\frac{1}{2}%
\left( \sqrt{12\pi }\varphi \right) ^{-2}+O\left( \left( \sqrt{12\pi }%
\varphi \right) ^{-3}\right) \right) .  \label{iatrcor}
\end{equation}
\end{description}

The corrections to (\ref{iatt1}) become of order the leading term when $%
\varphi \sim O\left( 1\right) ,$ that is, when the scalar field value drops
to the Planckian value or, more precisely, to $\varphi \simeq 1/\sqrt{12\pi }%
\simeq 1/6$. Hence solution (\ref{isol3}) is a good approximation only when
the scalar field exceeds the Planckian value. This does not mean, however,
that we require a theory of nonperturbative quantum gravity. Nonperturbative
quantum gravity effects become relevant only if the curvature or the energy
density reach the Planckian values. However, even for very large values of
the scalar field they can still remain in the sub-Planckian domain. In fact,
considering a massive homogeneous field with negligible kinetic energy we
infer that the energy density reaches the Planckian value for $\varphi
\simeq m^{-1}.$ Therefore, if $m\ll 1,$ then for \ $m^{-1}>\varphi >1$ we
can safely disregard nonperturbative quantum gravity effects.

According to (\ref{isol3}) the scalar field decreases linearly with time
after joining the attractor. During the inflationary stage
\begin{equation*}
p\simeq -\varepsilon +m^{2}/12\pi .
\end{equation*}%
So when the potential energy density $\sim m^{2}\varphi ^{2},$ which
dominates the total energy density, drops to $m^{2},$ inflation is over$.$
At this time the scalar field is of order unity (in Planckian units).

Let us determine the time dependence of the scale factor during inflation.
Substituting (\ref{isol3}) into (\ref{ifreq}) and neglecting the kinetic
term, we obtain a simple equation which is readily integrated to yield
\begin{equation}
a\left( t\right) \simeq a_{f}\exp \left( -\frac{m^{2}}{6}\left(
t_{f}-t\right) ^{2}\right) \simeq a_{i}\exp \left( \frac{\left(
H_{i}+H\left( t\right) \right) }{2}\left( t-t_{i}\right) \right) ,
\label{iscf1}
\end{equation}%
where $a_{i}$ and $H_{i}$ are the initial values of the scale factor and the
Hubble parameter. Note that the Hubble constant $H\left( t\right) \simeq
\sqrt{4\pi /3}m\varphi \left( t\right) $ also linearly decreases with time.
It follows from (\ref{isol3}) that inflation lasts for%
\begin{equation}
\Delta t\simeq t_{f}-t_{i}\simeq \sqrt{12\pi }\left( \varphi _{i}/m\right) .
\end{equation}%
During this time interval the scale factor increases
\begin{equation}
\frac{a_{f}}{a_{i}}\simeq \exp \left( 2\pi \varphi _{i}^{2}\right)
\label{iscf2}
\end{equation}%
times. The results obtained are in good agreement with the previous rough
estimates (\ref{idur}) and (\ref{i14a}). Inflation lasts more than 75-efolds
if the initial value of the scalar field, $\varphi _{i},$ is four times
larger than the Planckian value. To obtain an estimate for the largest
possible increase of the scale factor during inflation, let us consider a
scalar field of mass $10^{13}$ GeV$.$ The maximal possible value of the
scalar field for which we still remain in the sub-Planckian domain is $%
\varphi _{i}\sim 10^{6},$ and hence
\begin{equation}
\left( \frac{a_{f}}{a_{i}}\right) _{max}\sim \exp \left( 10^{12}\right) .
\label{iscf3}
\end{equation}%
Thus, the actual duration of the inflationary stage can massively exceed the
75 e-folds needed. In this case our universe would constitute only a very
tiny piece of an incredibly large homogeneous domain which originated from
one causal region. The other important feature of inflation is that the
Hubble constant decreases only by a factor $10^{-6}$, while the scale factor
grows by the tremendous amount given in (\ref{iscf3}), that is,
\begin{equation*}
\frac{H_{i}}{H_{f}}<<<\frac{a_{f}}{a_{i}}.
\end{equation*}

\textbf{Graceful exit and afterwards.} After the field drops below the
Planckian value it begins to oscillate. To determine the attractor behavior
in this regime we note that%
\begin{equation}
\dot{\varphi}^{2}+m^{2}\varphi ^{2}=\frac{3}{4\pi }H^{2}
\end{equation}%
and use the Hubble parameter $H$ and the angular variable $\theta ,$ defined
via
\begin{equation}
\dot{\varphi}=\sqrt{\frac{3}{4\pi }}H\sin \theta ,\text{ \ \ \ }m\varphi =%
\sqrt{\frac{3}{4\pi }}H\cos \theta ,\text{\ \ }  \label{invar}
\end{equation}%
as the new independent variables. It is convenient to replace equation (\ref%
{ifoeq}) by a system of two first order differential equations for $H$ and $%
\theta $:
\begin{eqnarray}
\dot{H} &=&-3H^{2}\sin ^{2}\theta ,  \label{ianv} \\
\dot{\theta} &=&-m-\frac{3}{2}H\sin 2\theta ,  \label{ianv1}
\end{eqnarray}%
where a dot denotes the derivative with respect to physical time $t.$ The
second term on the right hand side in (\ref{ianv1}) describes oscillations
with \textit{decaying} amplitude, as is evident from (\ref{ianv}).
Therefore, neglecting this term we obtain
\begin{equation}
\theta \simeq -mt+\alpha ,  \label{ianv22}
\end{equation}%
where the constant phase $\alpha $ can be set to zero. Thus, the scalar
field oscillates with frequency $\omega \simeq m.$ After substituting $%
\theta \simeq -mt$ into (\ref{ianv}), we obtain a readily integrated
equation with solution
\begin{equation}
H\left( t\right) \equiv \left( \frac{\dot{a}}{a}\right) \simeq \frac{2}{3t}%
\left( 1-\frac{\sin \left( 2mt\right) }{2mt}\right) ^{-1},  \label{isolr}
\end{equation}%
where a constant of integration is removed by a time shift. This solution is
applicable only for $mt\gg 1$. Therefore the oscillating term is small
compared to unity and the expression on the right hand side in (\ref{isolr})
can be expanded in powers of $\left( mt\right) ^{-1}.$ Substituting (\ref%
{ianv22}) and (\ref{isolr}) into the second equation in (\ref{invar}), we
obtain
\begin{equation}
\varphi \left( t\right) \simeq \frac{\cos \left( mt\right) }{\sqrt{3\pi }mt}%
\left( 1+\frac{\sin \left( 2mt\right) }{2mt}\right) +O\left( \left(
mt\right) ^{-3}\right) .  \label{iscf31}
\end{equation}%
The time dependence of the scale factor can easily be derived by integrating
(\ref{isolr}):
\begin{equation}
a\propto t^{2/3}\left( 1-\frac{\cos \left( 2mt\right) }{6m^{2}t^{2}}-\frac{1%
}{24m^{2}t^{2}}+O\left( \left( mt\right) ^{-3}\right) \right) .  \label{iscf}
\end{equation}%
Thus, in the leading approximation (up to decaying oscillating corrections),
the universe expands like a mat\-ter-dom\-i\-nat\-ed universe with zero
pressure. This is not surprising because an oscillating homogeneous field
can be thought of as a condensate of massive scalar particles with zero
momenta. Although the oscillating corrections are completely negligible in
the expressions for$\ a\left( t\right) $ and $H\left( t\right) ,$ they must
nevertheless be taken into account when we calculate the curvature
invariants. For example, the scalar curvature is
\begin{equation}
R\simeq -\frac{4}{3t^{2}}\left( 1+3\cos \left( 2mt\right) +O\left( \left(
mt\right) ^{-1}\right) \right)   \label{icur}
\end{equation}%
(compare to $R=-4/3t^{2}$ in a mat\-ter-dom\-i\-nat\-ed universe).

We have shown that inflation with a smooth graceful exit occurs naturally in
models with classical massive scalar fields. If the mass is small compared
to the Planck mass, the inflationary stage lasts long enough and is followed
by a cold mat\-ter-dom\-i\-nat\-ed stage. This cold matter, consisting of
heavy scalar particles, must finally be converted to radiation, baryons and
leptons. We will see later that this can easily be achieved in a variety of
ways.

\subsection{General potential: slow-roll approximation}

Equation (\ref{iscfeq}) for a massive scalar field in an expanding universe
coincides with the equation for a harmonic oscillator with a friction term
proportional to the Hubble parameter $H.$ It is well known that a large
friction damps the initial velocities and enforces a slow-roll regime in
which the acceleration can be neglected compared to the friction term.
Because for a general potential $H\propto \sqrt{\varepsilon }\sim \sqrt{V},$
we expect that for large values of $V$ the friction term can also lead to a
slow-roll inflationary stage, where $\ddot{\varphi}$ is negligible compared
to $3H\dot{\varphi}$. Omitting the $\ddot{\varphi}$ term and assuming that $%
\dot{\varphi}^{2}\ll V,$ equations (\ref{iscfeq}) and (\ref{ifreq}) simplify
to
\begin{equation}
3H\dot{\varphi}+V_{,\varphi }\simeq 0,\text{ \ \ \ \ }H\equiv \left( \frac{%
d\ln a}{dt}\right) \simeq \sqrt{\frac{8\pi }{3}V\left( \varphi \right) }.
\label{ieqslr}
\end{equation}%
Taking into account that
\begin{equation*}
\frac{d\ln a}{dt}=\dot{\varphi}\frac{d\ln a}{d\varphi }\simeq -\frac{%
V_{,\varphi }}{3H}\frac{d\ln a}{d\varphi },
\end{equation*}%
equations (\ref{ieqslr}) give
\begin{equation}
-V_{,\varphi }\frac{d\ln a}{d\varphi }\simeq 8\pi V  \label{ic3}
\end{equation}%
and hence
\begin{equation}
a\left( \varphi \right) \simeq a_{i}\exp \left( 8\pi \int_{\varphi
}^{\varphi _{i}}\frac{V}{V_{,\varphi }}d\varphi \right) .  \label{islrsol}
\end{equation}%
This approximate solution is valid only if the slow-roll conditions
\begin{equation}
\left\vert \dot{\varphi}^{2}\right\vert \ll \left\vert V\right\vert ,\text{
\ \ \ \ \ \ }\left\vert \ddot{\varphi}\right\vert \ll 3H\dot{\varphi}\sim
\left\vert V_{,\varphi }\right\vert ,  \label{ic5}
\end{equation}%
used to simplify equations (\ref{iscfeq}) and (\ref{ifreq}), are satisfied.
With the help of equations (\ref{ieqslr}), they can easily be recast in
terms of requirements on the derivatives of the potential itself:
\begin{equation}
\left( \frac{V_{,\varphi }}{V}\right) ^{2}\ll 1,\text{ \ \ \ \ \ }\left\vert
\frac{V_{,\varphi \varphi }}{V}\right\vert \ll 1.  \label{islrcond}
\end{equation}%
For a power-law potential, $V=\left( 1/n\right) \lambda \varphi ^{n},$ both
conditions are satisfied for $\left\vert \varphi \right\vert \gg 1.$ In this
case the scale factor changes as
\begin{equation}
a\left( \varphi \left( t\right) \right) \simeq a_{i}\exp \left( \frac{4\pi }{%
n}\left( \varphi _{i}^{2}-\varphi ^{2}\left( t\right) \right) \right) .
\label{iplpot}
\end{equation}%
It is obvious that the bulk\ of the inflationary expansion takes place when
the scalar field decreases by a factor of a few from its initial value.
However, we are interested mainly in the last 50-70 e-folds of inflation
because they determine the structure of the universe on present observable
scales. The detailed picture of the expansion during these last 70 e-folds
depends on the shape of the potential only within a rather narrow interval
of scalar field values.

\begin{description}
\item \addtocounter{prob}{+1} \textbf{Problem \theprob. }Find the time
dependence of the scale factor for the power law potential and estimate the
duration of inflation.

\item \addtocounter{prob}{+1} \textbf{Problem \theprob. }Verify that for a
general potential $V$ the system of equations (\ref{iscfeq}), (\ref{ifreq})
can be reduced to the following first order differential equation:
\begin{equation}
\frac{dy}{dx}=-3\left( 1-y^{2}\right) \left( 1+\frac{V_{,x}}{6yV}\right) ,
\label{ipr1}
\end{equation}%
where
\begin{equation*}
x\equiv \sqrt{\frac{4\pi }{3}}\varphi ;\text{ \ \ }y\equiv \sqrt{\frac{4\pi
}{3}}\frac{d\varphi }{d\ln a}.
\end{equation*}%
Assuming that $V_{,\varphi }/V\rightarrow 0$ as $\left\vert \varphi
\right\vert \rightarrow \infty ,$ draw the phase diagram and analyze the
behavior of the solutions in different asymptotic regions. Consider
separately the case of the exponential potential. What is the physical
meaning of the solutions in the regions corresponding to $y>1?$
\end{description}

After the end of inflation the scalar field begins to oscillate and the
universe enters the stage of deceleration. Assuming that the period of
oscillation is smaller than the cosmological time, let us determine the
effective equation of state. Neglecting the expansion and multiplying
equation (\ref{iscfeq}) by $\varphi ,$ we obtain
\begin{equation}
\left( \varphi \dot{\varphi}\right) ^{\cdot }-\dot{\varphi}^{2}+\varphi
V_{,\varphi }\simeq 0.  \label{iscosc}
\end{equation}%
As a result of averaging over a period, the first term vanishes and hence $%
\left\langle \dot{\varphi}^{2}\right\rangle \simeq \left\langle \varphi
V_{,\varphi }\right\rangle .$ Thus, the averaged effective equation of state
for an oscillating scalar field is

\begin{figure}
\begin{center}
\psfrag{x}[tr]{\small$\varphi$}
\psfrag{V}[tr]{\small$V$}
\psfrag{fc}[t]{\small\ $ \varphi_{c}$}
\psfrag{F>fc}[bl]{\small$\vert\varphi\vert\gggtr\varphi_{c}$}
\psfrag{LNA}[tl]{\small$V\sim ln\frac{\vert\varphi\vert}{ \varphi_{c}}$}
\includegraphics[width=2in,angle=0]{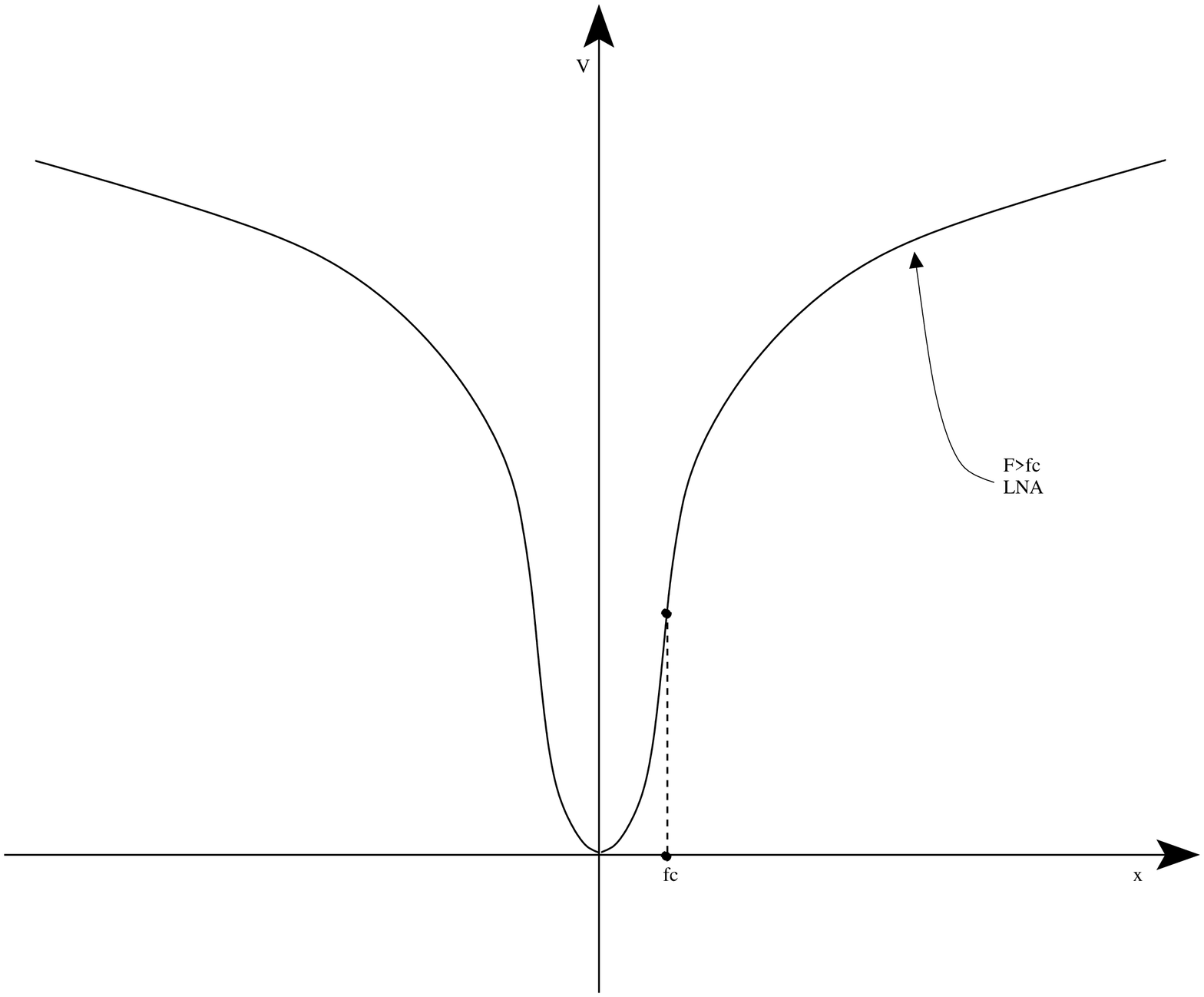}
\caption{~}
\end{center}
\end{figure}
\begin{equation}
w\equiv \frac{p}{\varepsilon }\simeq \frac{\left\langle \varphi V_{,\varphi
}\right\rangle -\left\langle 2V\right\rangle }{\left\langle \varphi
V_{,\varphi }\right\rangle +\left\langle 2V\right\rangle }.  \label{ieqst}
\end{equation}%
It follows that for $V\propto \varphi ^{n}$ we have $w\simeq \left(
n-2\right) /\left( n+2\right) .$ For an oscillating massive field $\left(
n=2\right) $ we obtain $w\simeq 0$ in agreement with our previous result. In
the case of a quartic potential $\left( n=4\right) ,$ the oscillating field
mimics an ultra-rel\-a\-tiv\-is\-tic fluid with $w\simeq 1/3.$

In fact, inflation can continue even after the end of slow-roll. Considering
the potential which behaves as
\begin{equation*}
V\sim \ln \left( \left\vert \varphi \right\vert /\varphi _{c}\right)
\end{equation*}%
for $1>\left\vert \varphi \right\vert \gg \varphi _{c}$ (see Fig.~4)$,$ we
infer from (\ref{ieqst}) that $w\rightarrow -1.$ This is easy to understand.
In the case of a convex potential, an oscillating scalar field spends most
of the time near the potential walls where its kinetic energy is negligible
and hence the main contribution to the equation of state comes from the
potential term.

\begin{description}
\item \addtocounter{prob}{+1} \textbf{Problem \theprob. }Which general
conditions must a potential $V$ satisfy to provide a stage of fast
oscillating inflation? How long can such inflation last and why is it not
very helpful for solving the initial conditions problem?
\end{description}

\section{Preheating and reheating}

The theory of reheating is far from complete. Not only the details, but even
the overall picture of inflaton decay depend crucially on the underlying
particle physics theory beyond the Standard Model. Because there are so many
possible extensions of the Standard Model, it does not make much sense to
study the particulars of the reheating processes in each concrete model.
Fortunately we are interested only in the final outcome of reheating,
namely, in the possibility of obtaining a thermal Friedmann universe.
Therefore, to illustrate the physical processes which could play a major
role we consider only simple toy\ models. The relative importance of the
different reheating mechanisms cannot be clarified without an underlying
particle theory. However, we will show that all of them lead to the desired
result.

\subsection{Elementary theory}

We consider an inflaton field $\varphi $ of mass $m$ coupled to a scalar
field $\chi $ and a spinor field $\psi .$ Their simplest interactions are
described by three leg diagrams (Fig.~5), which correspond to the following
terms in the Lagrangian:

\suppressfloats[t]
\begin{figure}
\begin{center}
\psfrag{1}{\small $\phi$} 
 \psfrag{3}{\small $\phi$}
\psfrag{4}[Bc][Bl]{\small $\bar{\psi}$}                                                
\psfrag{5}[Bc][Bl]{\small $\psi$}
\psfrag{a}[Bc][Bl]{\small $\chi$} 
\psfrag{b}[Bc][Bl]{\small $\chi$}
\includegraphics[width=11cm,angle=0]{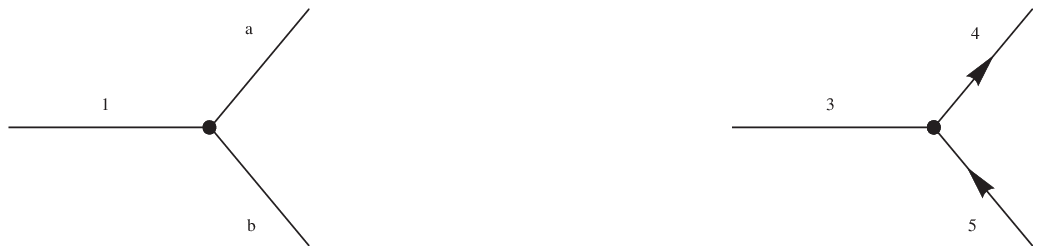}
\caption{~}\end{center}
\end{figure}
\begin{equation}
\Delta L_{int}=-g\varphi \chi ^{2}-h\varphi \bar{\psi}\psi .  \label{iintl}
\end{equation}%
We have seen that these kinds of couplings naturally arise in gauge theories
with spontaneously broken symmetry, and they are enough for our illustrative
purposes. To avoid a tachyonic instability we assume that $\left\vert
g\varphi \right\vert $ is smaller than the squared \textquotedblleft
bare\textquotedblright\ mass $m_{\chi }^{2}$. The decay rates of the
inflaton field into $\chi \chi -$ and $\bar{\psi}\psi -$pairs are determined
by the coupling constants $g$ and $h$ respectively$.$ They can easily be
calculated and the corresponding results are cited in every book on particle
physics:
\begin{equation}
\Gamma _{\chi }\equiv \Gamma \left( \varphi \rightarrow \chi \chi \right) =%
\frac{g^{2}}{8\pi m},\text{ \ \ \ }\Gamma _{\psi }\equiv \Gamma \left(
\varphi \rightarrow \psi \psi \right) =\frac{h^{2}m}{8\pi }.\text{\ \ }
\label{irates}
\end{equation}%
Let us apply these results in order to calculate the decay rate of the
inflaton. As we have noted, an \textit{oscillating }homogeneous scalar field
can be interpreted as a condensate of heavy particles of mass $m$
\textquotedblleft at rest,\textquotedblright\ that is, their 3-momenta $%
\mathbf{k}$ are equal to zero$.$ Keeping only the leading term in (\ref%
{iscf31}), we have
\begin{equation}
\varphi \left( t\right) \simeq \Phi \left( t\right) \cos \left( mt\right) ,
\label{iscfn1}
\end{equation}%
where $\Phi \left( t\right) $ is the slowly decaying amplitude of
oscillations. The number density of $\varphi $-particles can be estimated as
\begin{equation}
n_{\varphi }=\frac{\varepsilon _{\varphi }}{m}=\frac{1}{2m}\left( \dot{%
\varphi}^{2}+m^{2}\varphi ^{2}\right) \simeq \frac{1}{2}m\Phi ^{2}.
\label{ind}
\end{equation}%
This number is very large. For example, for $m\sim 10^{13}$ GeV$,$ we have $%
n_{\varphi }\sim 10^{92}$ cm$^{-3}$ immediately after the end of inflation,
when $\Phi \sim 1$ in Planckian units.

One can show that quantum corrections do not significantly modify the
interactions (\ref{iintl}) only if $g<m$ and $h<m^{1/2}$. Therefore, for $%
m\ll m_{Pl},$ the highest decay rate into $\chi -$particles, $\Gamma _{\chi
}\sim m,$ is much larger than the highest possible rate for the decay into
fermions, $\Gamma _{\psi }\sim m^{2}$. If $g\sim m,$ then the lifetime of a $%
\varphi -$particle is about $\Gamma _{\chi }^{-1}\sim m^{-1}$ and the
inflaton decays after a few oscillations. Even if the coupling is not so
large, the decay can still be very efficient. The reason is that the
effective decay rate into bosons, $\Gamma _{\text{eff}},$ is equal to $%
\Gamma _{\chi },$ given in (\ref{irates}), only if the phase space of $\chi
- $particles is not densely populated by previously created $\chi -$%
particles. Otherwise $\Gamma _{\text{eff}}$ can be made much larger by the
effect of Bose condensation. This amplification of the inflaton decay is
discussed in the next section.

Taking into account the expansion of the universe, the equations for the
number densities of the $\varphi -$ and $\chi -$particles can be written as
\begin{equation}
\frac{1}{a^{3}}\frac{d\left( a^{3}n_{\varphi }\right) }{dt}=-\Gamma _{\text{%
eff}}n_{\varphi };\text{ \ \ }\frac{1}{a^{3}}\frac{d\left( a^{3}n_{\chi
}\right) }{dt}\text{\ }=2\Gamma _{\text{eff}}n_{\varphi },  \label{ikin}
\end{equation}%
where the coefficient two in the second equation arises because \textit{one}
$\varphi -$particle decays into \textit{two} $\chi -$particles.

\begin{description}
\item \addtocounter{prob}{+1} \textbf{Problem \theprob. }Substituting (\ref%
{ind}) into the first equation in (\ref{ikin}), derive the \textit{%
approximate} equation
\begin{equation}
\ddot{\varphi}+\left( 3H+\Gamma _{\text{eff}}\right) \dot{\varphi}%
+m^{2}\varphi \simeq 0,  \label{ieffeq}
\end{equation}%
which shows that the decay of the inflaton amplitude due to particle
production may be roughly taken into account by introducing an extra
friction term $\Gamma _{\text{eff}}\dot{\varphi}.$ Why is this equation
applicable only during the oscillatory phase?
\end{description}

\subsection{Narrow resonance}

The domain of applicability of elementary reheating theory is limited. Bose
condensation effects become important very soon after the beginning of the
inflaton decay. Because the inflaton particle is \textquotedblleft at
rest,\textquotedblright\ the momenta of the two produced $\chi -$particles
have the same magnitude $k$ but opposite directions. If the corresponding
states in the phase space of $\chi -$particles are already occupied, then
the inflaton decay rate is enhanced by a bose factor. The inverse decay
process $\chi \chi \rightarrow \varphi $ can also take place. The rates of
these processes are proportional to
\begin{equation*}
\left\vert \left\langle n_{\varphi }-1,n_{\mathbf{k}}+1,n_{-\mathbf{k}%
}+1\right\vert \hat{a}_{\mathbf{k}}^{+}\hat{a}_{-\mathbf{k}}^{+}\hat{a}%
_{\varphi }^{-}\left\vert n_{\varphi },n_{\mathbf{k}},n_{-\mathbf{k}%
}\right\rangle \right\vert ^{2}=\left( n_{\mathbf{k}}+1\right) \left( n_{-%
\mathbf{k}}+1\right) n_{\varphi }
\end{equation*}%
and
\begin{equation*}
\left\vert \left\langle n_{\varphi }+1,n_{\mathbf{k}}-1,n_{-\mathbf{k}%
}-1\right\vert \hat{a}_{\varphi }^{+}\hat{a}_{\mathbf{k}}^{-}\hat{a}_{-%
\mathbf{k}}^{-}\left\vert n_{\varphi },n_{\mathbf{k}},n_{-\mathbf{k}%
}\right\rangle \right\vert ^{2}=n_{\mathbf{k}}n_{-\mathbf{k}}\left(
n_{\varphi }+1\right)
\end{equation*}%
respectively, where $\hat{a}_{\mathbf{k}}^{\pm }$ are the creation and
annihilation operators for $\chi -$ particles and $n_{\pm \mathbf{k}}$ are
their occupation numbers. To avoid confusion the reader must always
distinguish the occupation numbers from the number densities keeping in mind
that the occupation number refers to a density per cell of volume $\left(
2\pi \right) ^{3}$ (in the Planckian units) in the phase space, while the
number density is the number of particles per unit volume in the
three-dimensional space. Taking into account that $n_{\mathbf{k}}=n_{-%
\mathbf{k}}\equiv n_{k}$ and $n_{\varphi }\gg 1,$ we infer that the number
densities $n_{\varphi }$ and $n_{\chi }$ satisfy (\ref{ikin}), where
\begin{equation}
\Gamma _{\text{eff}}\simeq \Gamma _{\chi }(1+2n_{k}).  \label{ieff}
\end{equation}

\begin{figure}
\begin{center}
\psfrag{A}[b ]{\small$\Delta k\simeq \left( \frac{4g\Phi }{m}\right)$}
\psfrag{B}[Bl]{\small$k_{0}\simeq \frac{m}{2}$}
\psfrag{c}[b ]{\small$k_{*}/\sqrt{\pi}$}
\psfrag{d}[t]{\small a)}
\psfrag{e}[t]{\small b)}
\includegraphics[width=0.9\textwidth,angle=0]{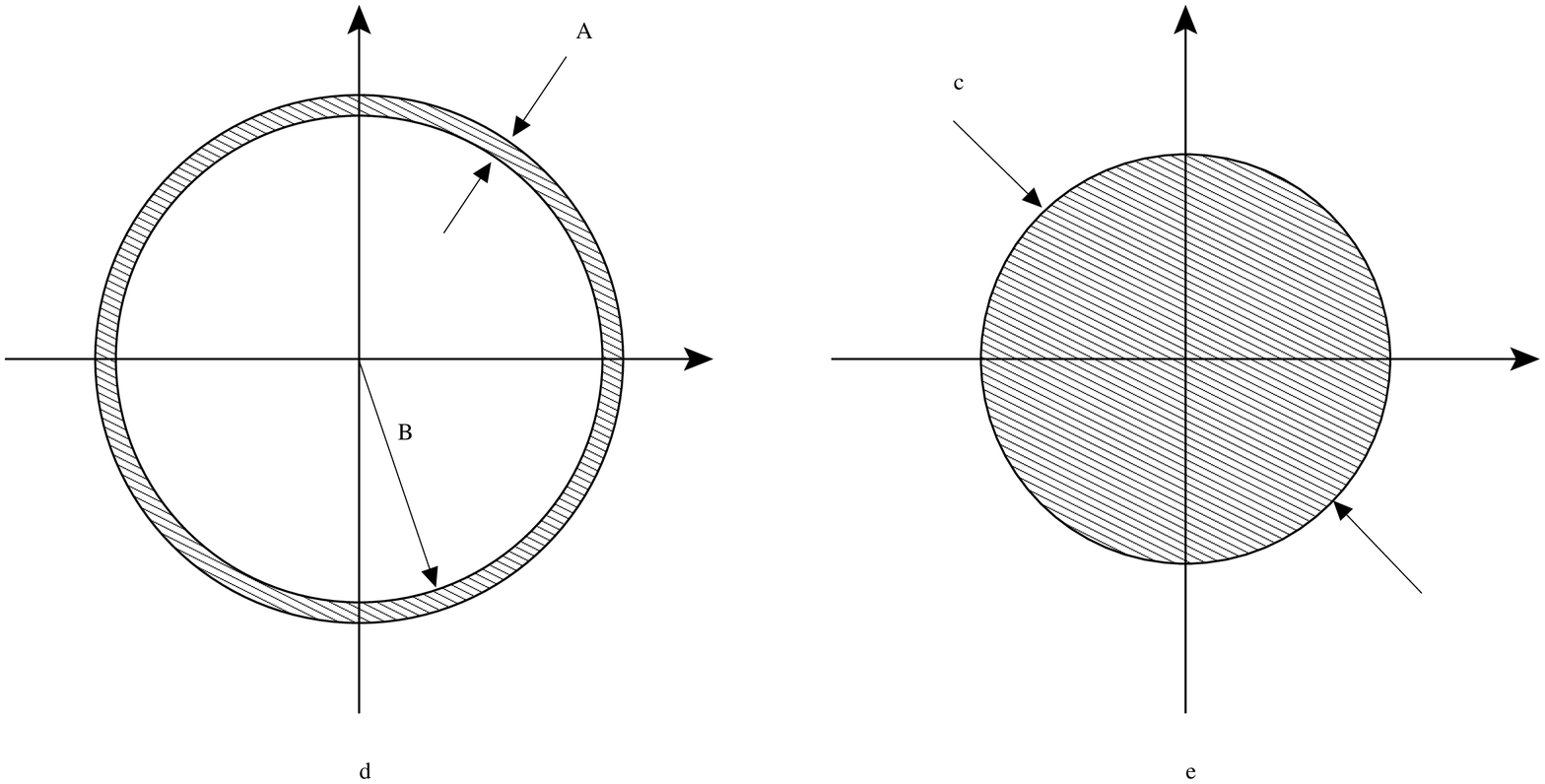}
\caption{~}
\end{center}
\end{figure}

Given a number density $n_{\chi },$ let us calculate $n_{k}.$ A $\varphi $%
-particle \textquotedblleft at rest\textquotedblright\ decays into two $\chi
-$particles, both having energy $m/2.$ Because of the interaction term (\ref%
{iintl}), the effective squared mass of the $\chi -$particle depends on the
value of the inflaton field and is equal to $m_{\chi }^{2}+2g\varphi \left(
t\right) .$ Therefore the corresponding 3-momentum of the produced $\chi -$%
particle is given by
\begin{equation}
k=\left( \left( \frac{m}{2}\right) ^{2}-m_{\chi }^{2}-2g\varphi \left(
t\right) \right) ^{1/2},  \label{i3mom}
\end{equation}%
where we assume that $m_{\chi }^{2}+2g\varphi \ll m^{2}.$ The oscillating
term,
\begin{equation*}
g\varphi \simeq g\Phi \cos \left( mt\right) ,
\end{equation*}%
leads to a \textquotedblleft scattering\textquotedblright\ of the momenta in
phase space. If $g\Phi \ll m^{2}/8,$ then all particles are created within a
thin shell of width
\begin{equation}
\Delta k\simeq m\left( \frac{4g\Phi }{m^{2}}\right) \ll m  \label{iwidth}
\end{equation}%
located near the radius $k_{0}\simeq m/2$ (Fig.~6a). Therefore
\begin{equation}
n_{k=m/2}\simeq \frac{n_{\chi }}{\left( 4\pi k_{0}^{2}\Delta k\right)
/\left( 2\pi \right) ^{3}}\simeq \frac{2\pi ^{2}n_{\chi }}{mg\Phi }=\frac{%
\pi ^{2}\Phi }{g}\frac{n_{\chi }}{n_{\varphi }}.  \label{ioccn}
\end{equation}%
The occupation numbers $n_{k}$ exceed unity, and hence, the Bose
condensation effect is essential only if
\begin{equation}
n_{\chi }>\frac{g}{\pi ^{2}\Phi }n_{\varphi }.  \label{ioccn1}
\end{equation}%
Taking into account that at the end of inflation $\Phi \sim 1,$ we infer
that the occupation numbers begin to exceed unity as soon as the inflaton
converts a fraction $g$ of its energy to $\chi -$particles. The derivation
above is valid only for $g\Phi \ll m^{2}/8$. Therefore, if $m\sim 10^{-6},$
then at most a fraction $g\sim m^{2}\sim 10^{-12}$ of the inflaton energy
can be transferred to $\chi -$particles in the regime where $n_{k}<1.$ Thus,
the elementary theory of reheating, which is applicable for $n_{k}\ll 1,$
fails almost immediately after the beginning of reheating. Given the result
in (\ref{ioccn}), the effective decay rate (\ref{ieff}) becomes
\begin{equation}
\Gamma _{\text{eff}}\simeq \frac{g^{2}}{8\pi m}\left( 1+\frac{2\pi ^{2}\Phi
}{g}\frac{n_{\chi }}{n_{\varphi }}\right) ,  \label{ioccn2}
\end{equation}%
where we have used equation (\ref{irates}) for $\Gamma _{\chi }.$
Substituting this expression into the second equation in (\ref{ikin}), we
obtain
\begin{equation}
\frac{1}{a^{3}}\frac{d\left( a^{3}n_{\chi }\right) }{dN}\text{\ }=\frac{g^{2}%
}{2m^{2}}\left( 1+\frac{2\pi ^{2}\Phi }{g}\frac{n_{\chi }}{n_{\varphi }}%
\right) n_{\varphi },  \label{ieqmod}
\end{equation}%
where $N\equiv mt/2\pi $ is the number of inflaton oscillations. Let us
neglect for a moment the expansion of the universe and disregard the
decrease of the inflation amplitude due to particle production. In this case
$\Phi =const$ and for $n_{k}\gg 1$ equation (\ref{ieqmod}) can be easily
integrated$.$ The result is
\begin{equation}
n_{\chi }\propto \exp \left( \frac{\pi ^{2}g\Phi }{m^{2}}N\right) \propto
\exp \left( 2\pi \mu N\right) ,\text{ \ }  \label{iexpgr}
\end{equation}%
where $\mu \equiv \pi g\Phi /\left( 2m^{2}\right) $ is the parameter of
instability.

\begin{description}
\item \addtocounter{prob}{+1} \textbf{Problem \theprob. }Derive the
following equation for the Fourier modes of the field $\chi $ in Minkowski
space:
\begin{equation}
\ddot{\chi}_{k}+\left( k^{2}+m_{\chi }^{2}+2g\Phi \cos mt\right) \chi _{k}=0.
\label{ioscil}
\end{equation}%
Reduce it to the well-known Mathieu equation and, assuming that $m^{2}\gg
m_{\chi }^{2}\geq 2\left\vert g\Phi \right\vert ,$ investigate the narrow
parametric resonance. Determine the instability bands and the corresponding
instability parameters. Compare the width of the first instability band with
(\ref{iwidth}). Where is this band located? The minimal value of the initial
amplitude of $\chi _{k}$ is due to vacuum fluctuations. The increase of $%
\chi _{k}$ with time can be interpreted as the production of $\chi -$%
particles by the external classical field $\varphi ,$ with $n_{\chi }\propto
$ $\left\vert \chi _{k}\right\vert ^{2}.$ Show that in the center of the
first instability band,
\begin{equation}
n_{\chi }\propto \exp \left( \frac{4\pi g\Phi }{m^{2}}N\right) ,
\label{iincr}
\end{equation}%
where $N$ is the number of oscillations. Compare this result with (\ref%
{iexpgr}) and explain why they are different by a numerical factor in the
exponent. Thus, Bose condensation can be interpreted as a narrow parametric
resonance in the first instability band, and vice versa. Give a physical
interpretation of the higher order resonance bands in terms of particle
production.
\end{description}

Using the results of this problem we can reduce the investigation of the
inflaton decay due to the coupling

\begin{equation}
\Delta L_{int}=-\frac{1}{2}\tilde{g}^{2}\varphi ^{2}\chi ^{2},
\label{iquart}
\end{equation}%
to the case studied above. In fact, the equation for a massless scalar field
$\chi $, coupled to the inflaton $\varphi =\Phi \cos mt,$ takes the form
\begin{equation}
\ddot{\chi}_{k}+\left( k^{2}+\tilde{g}^{2}\Phi ^{2}\cos ^{2}mt\right) \chi
_{k}=0,  \label{iquart1}
\end{equation}%
which coincides with equation (\ref{ioscil}) for $m_{\chi }^{2}=2g\Phi $
after substitutions $\tilde{g}^{2}\Phi ^{2}\rightarrow 4g\Phi $ and $%
m\rightarrow m/2.$ Thus, the two problems are mathematically equivalent.
Using this observation and making the corresponding replacements in formula (%
\ref{iincr}), we immediately find that
\begin{equation}
n_{\chi }\propto \exp \left( \frac{\pi \tilde{g}^{2}\Phi ^{2}}{4m^{2}}%
N\right) .  \label{iquart2}
\end{equation}%
The condition for narrow resonance is $\tilde{g}\Phi \ll m$ and the width of
the first resonance band can be estimated from (\ref{iwidth}) as $\Delta
k\sim m\left( \tilde{g}^{2}\Phi ^{2}/m^{2}\right) .$

In summary, we have shown that even for a small coupling constant the
elementary theory of reheating must be modified to take into account the
Bose condensation effect, and that this can lead to an exponential increase
of the reheating efficiency.

\begin{description}
\item \addtocounter{prob}{+1} \textbf{Problem \theprob. }Taking a few
concrete values for $g$ and $m,$ compare the results of the elementary
theory with those obtained for narrow parametric resonance.
\end{description}

So far we have neglected the expansion of the universe, the backreaction of
the produced particles and their rescatterings. All these effects work to
suppress the efficiency of the narrow parametric resonance. The expansion
shifts the momenta of the previously created particles and takes them out of
the resonance layer (Fig.~6a). Thus, the occupation numbers relevant for
Bose condensation are actually smaller than what one would expect according
to the naive estimate (\ref{ioccn}). If the rate of supply of newly created
particles in the resonance layer is smaller than the rate of their escape,
then $n_{k}<1$ and we can use the elementary theory of reheating. The other
important effect is the decrease of the amplitude $\Phi \left( t\right) $
due to both the expansion of the universe and particle production. Because
the width of the resonance layer is proportional to $\Phi ,$ it becomes more
and more narrow. As a result the particles can escape from this layer more
easily and they do not stimulate the subsequent production of particles. The
rescattering of the $\chi -$particles also suppresses the resonance
efficiency by removing particles from the resonance layer. Another effect is
the change of the effective inflaton mass due to the newly produced $\chi -$%
particles; this shifts the center of the resonance layer from its original
location.

To conclude, narrow parametric resonance is very sensitive to the interplay
of different complicating factors. It can be fully investigated only using
numerical methods. From our analytical consideration we can only say that
the inflaton field probably decays not as \textquotedblleft
slowly\textquotedblright\ as in the elementary theory, but not as
\textquotedblleft fast\textquotedblright\ as in the case of pure narrow
parametric resonance.

\subsection{Broad resonance}

\index{broad resonance}So far we have considered only the case of a small
coupling constants. Quantum corrections to the Lagrangian are not very
crucial if $g<m$ and $%
\tilde{g}<\left( m/\Phi \right) ^{1/2}.$ They can therefore be ignored when
we consider inflaton decay in the strong coupling regime: $m>g>m^{2}/\Phi $
for the three-leg interaction and $\left( m/\Phi \right) ^{1/2}>\tilde{g}%
>m/\Phi $ for the quartic interaction (\ref{iquart}). In this case the
condition for narrow resonance is not fulfilled and we cannot use the
methods above. Perturbative methods fail because the higher order diagrams,
built from the elementary diagrams, give comparable contributions. Particle
production can be treated only as a collective effect in which many inflaton
particles participate simultaneously. We have to apply the methods of
quantum field theory in an external classical background $-$ as in Problem
1.11.

Let us consider quartic interaction (\ref{iquart}). First, we neglect the
expansion of the universe. For $\tilde{g}\Phi \gg m$ the mode equation (see (%
\ref{iquart1})):
\begin{equation}
\ddot{\chi}_{k}+\omega ^{2}\left( t\right) \chi _{k}=0,  \label{ibrres}
\end{equation}%
where
\begin{equation}
\omega \left( t\right) \equiv \left( k^{2}+\tilde{g}^{2}\Phi ^{2}\cos
^{2}mt\right) ^{1/2},  \label{ifr}
\end{equation}%
describes a broad parametric resonance. If the frequency $\omega \left(
t\right) $ is a slowly varying function of time or, more precisely, $%
\left\vert \dot{\omega}\right\vert \ll \omega ^{2},$ equation (\ref{ibrres})
can be solved in the quasiclassical (WKB) approximation:
\begin{equation}
\chi _{k}\propto \frac{1}{\sqrt{\omega }}\exp \left( \pm i\int \omega
dt\right) .  \label{iqclsol}
\end{equation}%
In this case the number of particles, $n_{\chi }\sim \varepsilon _{\chi
}/\omega ,$ is an adiabatic invariant and is conserved. For most of the time
the condition $\left\vert \dot{\omega}\right\vert \ll \omega ^{2}$ is indeed
fulfilled. However, every time the oscillating inflaton vanishes at $%
t_{j}=m^{-1}\left( j+1/2\right) \pi ,$ the effective mass of the $\chi $
field, proportional to $\left\vert \cos \left( mt\right) \right\vert ,$
vanishes. It is shortly before and after $t_{j}$ that the adiabatic
condition is strongly violated:
\begin{equation}
\frac{\left\vert \dot{\omega}\right\vert }{\omega ^{2}}=\frac{m\tilde{g}%
^{2}\Phi ^{2}\left\vert \cos \left( mt\right) \sin \left( mt\right)
\right\vert }{\left( k^{2}+\tilde{g}^{2}\Phi ^{2}\cos ^{2}\left( mt\right)
\right) ^{3/2}}\geq 1.
\end{equation}%
Considering a small time interval $\Delta t\ll m^{-1}$ in the vicinity of $%
t_{j},$ we can rewrite this condition as
\begin{equation}
\frac{\Delta t/\Delta t_{\ast }}{\left( k^{2}\Delta t_{\ast }^{2}+\left(
\Delta t/\Delta t_{\ast }\right) ^{2}\right) ^{3/2}}\geq 1,  \label{iadcond1}
\end{equation}%
where
\begin{equation}
\Delta t_{\ast }\simeq \left( \tilde{g}\Phi m\right) ^{-1/2}=\frac{1}{m}%
\left( \tilde{g}\Phi /m\right) ^{-1/2}.  \label{iint}
\end{equation}%
It follows that the adiabatic condition is broken only within short time
intervals $\Delta t\sim \Delta t_{\ast }$ near $t_{j}$ and only for modes
with
\begin{equation}
k<k_{\ast }\simeq \Delta t_{\ast }^{-1}\simeq m\left( \tilde{g}\Phi
/m\right) ^{1/2}.  \label{imom}
\end{equation}%
Therefore, we expect that $\chi -$particles with the corresponding momenta
are created only during these time intervals. It is worth noting that the
momentum of the created particle can be larger than the inflaton mass by the
ratio $\left( \tilde{g}\Phi /m^{2}\right) ^{1/2}>1;$ the $\chi -$particles
are produced as a result of a collective process involving many inflaton
particles. This is the reason why we cannot describe the broad resonance
regime using the usual methods of perturbation theory.

To calculate the number of particles produced in a single inflaton
oscillation we consider a short time interval in the vicinity of $t_{j}$ and
approximate the cosine in (\ref{ibrres}) by a linear function. Equation (\ref%
{ibrres}) then takes the form
\begin{equation}
\frac{d^{2}\chi _{\kappa }}{d\tau ^{2}}+\left( \kappa ^{2}+\tau ^{2}\right)
\chi _{\kappa }=0,  \label{iapp}
\end{equation}%
where the dimensionless wavenumber $\kappa \equiv k/k_{\ast }$ and time $%
\tau \equiv \left( t-t_{j}\right) /\Delta t_{\ast }$ have been introduced$.$
In terms of the new variables the adiabaticity condition is broken at $%
\left\vert \tau \right\vert <1$ and only for $\kappa <1.$ It is remarkable
that the coupling constant $\tilde{g}$, the mass and the amplitude of the
inflaton enter explicitly only in $\kappa ^{2}.$ The adiabaticity violation
is largest for $k=0.$ In this case the parameters $\tilde{g},\Phi $ and $m$
drop from equation (\ref{iapp}) and the amplitude $\chi _{\kappa =0}$
changes only by a numerical, parameter-independent factor as a result of
passing through the nonadiabatic region at $\left\vert \tau \right\vert <1.$
Because the particle density $n$ is proportional to $\left\vert \chi
\right\vert ^{2},$ its growth from one oscillation to the next can written
as
\begin{equation}
\left( \frac{n^{j+1}}{n^{j}}\right) _{k=0}=\exp \left( 2\pi \mu
_{k=0}\right) ,  \label{ichp}
\end{equation}%
where the instability parameter $\mu _{k=0}$ does not depend on $\tilde{g},$
$\Phi $ and $m.$ For modes with $k\neq 0,$ the parameter $\mu _{k\neq 0}$ is
a function of $\kappa =k/k_{\ast }.$ In this case the adiabaticity is not
violated as strongly as for the $k=0$ mode, and hence $\mu _{k\neq 0}$ is
smaller than $\mu _{k=0}.$ To calculate the instability parameters we have
to determine the change of the amplitude $\chi $ in passing from the $\tau
<-1$ region to the $\tau >1$ region$.$ This can be done using two
independent WKB solutions of equation (\ref{iapp}) in the asymptotic regions
$\left\vert \tau \right\vert \gg 1$:
\begin{equation}
\chi _{\pm }=\frac{1}{\left( \kappa ^{2}+\tau ^{2}\right) ^{1/4}}\exp \left(
\pm i\int \sqrt{\kappa ^{2}+\tau ^{2}}d\tau \right) \simeq \left\vert \tau
\right\vert ^{-\frac{1}{2}\pm \frac{1}{2}i\kappa ^{2}}\exp \left( \pm \frac{%
i\tau ^{2}}{2}\right) .  \label{iqcl}
\end{equation}%
After passing through the nonadiabatic region the mode $A_{+}\chi _{+}$
becomes a mixture of the modes $\chi _{+}$ and $\chi _{-}$ , that is,
\begin{equation}
A_{+}\chi _{+}\rightarrow B_{+}\chi _{+}+C_{+}\chi _{-},  \label{imtr}
\end{equation}%
where $A_{+},B_{+}$ and $C_{+}$ are the complex constant coefficients.
Similarly, for the mode $A_{-}\chi _{-},$ we have
\begin{equation}
A_{-}\chi _{-}\rightarrow B_{-}\chi _{-}+C_{-}\chi _{+}.  \label{imtr1}
\end{equation}%
Drawing an analogy with the scattering problem for the inverse parabolic
potential, we note that the mixture arises due to an overbarrier reflection
of the wave. The reflection is most efficient for the waves with $k=0$ which
\textquotedblleft touch\textquotedblright\ the top of the barrier.

The quasiclassical solution is valid in the complex plane for $\left\vert
\tau \right\vert \gg 1.$ Traversing the appropriate contour $\tau
=\left\vert \tau \right\vert e^{i\varphi }$ in the complex plane from $\tau
\ll -1$ to $\tau \gg 1,$ we infer from (\ref{iqcl}), (\ref{imtr}) and (\ref%
{imtr1}) that
\begin{equation}
B_{\pm }=\mp ie^{-\frac{\pi }{2}\kappa ^{2}}A_{\pm }.  \label{iba}
\end{equation}%
The coefficients $C_{\pm }$ are not determined in this method. To find them
we use the Wronskian%
\begin{equation}
W\equiv \dot{\chi}\chi ^{\ast }-\chi \dot{\chi}^{\ast },
\end{equation}%
where $\chi $ is an arbitrary complex solution of equation (\ref{iapp}).
Taking the derivative of $W$ and using (\ref{iapp}) to express $\ddot{\chi}$
in terms of $\chi ,$ we find%
\begin{equation}
\dot{W}=0
\end{equation}%
and hence $W=const.$ From this we infer that the coefficients $A,B$ and $C$
in (\ref{imtr}) and (\ref{imtr1}) satisfy the \textquotedblleft probability
conservation\textquotedblright\ condition

\begin{equation}
\left| C_{\pm }\right| ^{2}-\left| B_{\pm }\right| ^{2}=\left| A_{\pm
}\right| ^{2}.  \label{inorm}
\end{equation}
Substituting $B$ from (\ref{iba}), we obtain
\begin{equation}
C_{\pm }=\sqrt{1+e^{-\pi \kappa ^{2}}}\left| A_{\pm }\right| e^{i\alpha
_{\pm }},  \label{inorm1}
\end{equation}
where the phases $\alpha _{\pm }$ remain undetermined.

At $\left\vert \tau \right\vert \gg 1$ the modes of field $\chi $ satisfy
the harmonic oscillator equation with a slowly changing frequency $\omega
\propto \left\vert \tau \right\vert .$ In quantum field theory the
occupation number $n_{k}$ in the expression for the energy of the harmonic
oscillator,
\begin{equation}
\varepsilon _{k}=\omega \left( n_{k}+1/2\right) ,
\end{equation}%
is interpreted as the number of particles in the corresponding mode $k$.\ In
the adiabatic regime ($\left\vert \tau \right\vert \gg 1$) this number is
conserved and it changes only when the adiabatic condition is violated. Let
us consider an arbitrary initial mixture of the modes $\chi _{+}$ and $\chi
_{-}$. After passing through the nonadiabatic region at $t\sim t_{j},$ it
changes as
\begin{equation}
\chi ^{j}=A_{+}\chi _{+}+A_{-}\chi _{-}\rightarrow \chi ^{j+1}=\left(
B_{+}+C_{-}\right) \chi _{+}+\left( B_{-}+C_{+}\right) \chi _{-}.
\label{iar}
\end{equation}%
Taking into account that
\begin{equation}
n+\frac{1}{2}=\frac{\varepsilon }{\omega }\simeq \omega \left\vert \chi
\right\vert ^{2},
\end{equation}%
we see that as a result of this passage the number of particles in the mode $%
k$ increases
\begin{equation}
\left( \frac{n^{j+1}+1/2}{n^{j}+1/2}\right) _{k}\simeq \frac{\omega
\left\vert \chi ^{j+1}\right\vert ^{2}}{\omega \left\vert \chi
^{j}\right\vert ^{2}}\simeq \frac{\left\vert B_{+}+C_{-}\right\vert
^{2}+\left\vert B_{-}+C_{+}\right\vert ^{2}}{\left\vert A_{+}\right\vert
^{2}+\left\vert A_{-}\right\vert ^{2}}  \label{iar2}
\end{equation}%
times, where we have averaged $\left\vert \chi \right\vert ^{2}$ over the
time interval $m^{-1}>t>\omega ^{-1}$. With $B$ and $C$ from (\ref{iba}) and
(\ref{inorm1}), this expression becomes
\begin{equation}
\left( \frac{n^{j+1}+1/2}{n^{j}+1/2}\right) _{k}\simeq \left( 1+2e^{-\pi
\kappa ^{2}}\right) +\frac{4\left\vert A_{-}\right\vert \left\vert
A_{+}\right\vert }{\left\vert A_{+}\right\vert ^{2}+\left\vert
A_{-}\right\vert ^{2}}\cos \theta e^{-\frac{\pi }{2}\kappa ^{2}}\sqrt{%
1+e^{-\pi \kappa ^{2}}}.  \label{ienr}
\end{equation}

\begin{description}
\item \addtocounter{prob}{+1} \textbf{Problem \theprob. }Verify the result (%
\ref{ienr}) and explain the origin of the phase $\theta .$
\end{description}

In the vacuum initial state $n_{k}=0$ but the amplitude of the field $\chi $
does not vanish because of the existence of vacuum fluctuations; we have $%
\left\vert A_{+}^{0}\right\vert ^{2}\neq 0$ and $\left\vert
A_{-}^{0}\right\vert ^{2}=0.$ It follows from the \textquotedblleft
probability conservation\textquotedblright\ condition that
\begin{equation}
\left\vert A_{+}\right\vert ^{2}-\left\vert A_{-}\right\vert ^{2}=\left\vert
A_{+}^{0}\right\vert ^{2}  \label{ivnor}
\end{equation}%
at every moment of time. This means that as a result of particle production
the coefficients $\left\vert A_{+}\right\vert ^{2}$ and $\left\vert
A_{-}\right\vert ^{2}$ grow by the same amount. When $\left\vert
A_{+}\right\vert $ becomes much larger than $\left\vert A_{+}^{0}\right\vert
$ we have $\left\vert A_{+}\right\vert \simeq \left\vert A_{-}\right\vert .$
Taking this into account and beginning in the vacuum state, we find from (%
\ref{ienr}) that after $N\gg 1$ inflaton oscillations the particle number in
mode $k$ is
\begin{equation}
n_{k}\simeq \frac{1}{2}\exp \left( 2\pi \mu _{k}N\right) ,  \label{inos}
\end{equation}%
where the instability parameter is given by
\begin{equation}
\mu _{k}\simeq \frac{1}{2\pi }\ln \left( 1+2e^{-\pi \kappa ^{2}}+2\cos
\theta e^{-\frac{\pi }{2}\kappa ^{2}}\sqrt{1+e^{-\pi \kappa ^{2}}}\right) .
\label{iinc1}
\end{equation}%
This parameter takes its maximal value
\begin{equation*}
\mu _{k}^{\max }=\pi ^{-1}\ln \left( 1+\sqrt{2}\right) \simeq 0.28
\end{equation*}%
for $k=0$ and $\theta =0.$ In the interval $-\pi <\theta <\pi $ we find that
$\mu _{k=0}$ is positive if $3\pi /4>\theta >-3\pi /4$ and negative
otherwise. Thus, assuming random $\theta $, we conclude that the particle
number in every mode changes stochastically. However, if all $\theta $ are
equally probable, then the number of particles increases three quarters of
the time and therefore it also increases on average, in agreement with
entropic arguments. The net instability parameter, characterizing the
average growth in particle number, is obtained by skipping the $\cos \theta $%
-term in (\ref{iinc1}):
\begin{equation}
\bar{\mu}_{k}\simeq \frac{1}{2\pi }\ln \left( 1+2e^{-\pi \kappa ^{2}}\right)
.  \label{iavi}
\end{equation}

With slight modifications the results above can be applied to an expanding
universe. First of all we note that the expansion randomizes the phases $%
\theta $ and hence the effective instability parameter is given by (\ref%
{iavi}). For particles with physical momenta $k<k_{\ast }/\sqrt{\pi }$, the
instability parameter $\bar{\mu}_{k}$ can be roughly estimated by its value
at the center of the instability region, $\bar{\mu}_{k=0}=\ln 3/2\pi \simeq
0.175.$ To understand how the expansion can influence the efficiency of
broad resonance, it is again helpful to use the phase space picture. The
particles created in the broad resonance regime occupy the entire sphere of
radius $k_{\ast }/\sqrt{\pi }$ in phase space (see Fig. 6b). During the
passage through the nonadiabatic region the number of particles in every
cell of the sphere, and hence the total number density, increases on average
$\exp \left( 2\pi \cdot 0.175\right) \simeq 3$ times. At the stage when
inflaton energy is still dominant, the physical momentum of the created
particle decreases \ in inverse proportion to the scale factor $\left(
k\propto a^{-1}\right) $, while the radius of the sphere shrinks more
slowly, namely, as $\Phi ^{1/2}\propto t^{-1/2}\propto a^{-3/4}$. As a
result, the created particles move away from the boundary of the sphere
towards its center where they participate in the next \textquotedblleft act
of creation,\textquotedblright\ enhancing the probability by a bose factor.
Furthermore, expansion also makes broad resonance less sensitive to
rescattering and backreaction effects. These two effects influence the
resonance efficiency by removing those particles which are located near the
boundary of the resonance sphere. Because expansion moves particles away
from this\ region, the impact of these effects is diminished. Thus, in
contrast to the narrow resonance case, expansion stabilizes broad resonance
and at the beginning of reheating it can be realized in its pure form.

Taking into account that the initial volume of the resonance sphere is about
\begin{equation*}
k_{\ast }^{3}\simeq m^{3}\left( \tilde{g}\Phi _{0}/m\right) ^{3/2},
\end{equation*}%
we obtain the following estimate for the ratio of the particle number
densities after $N$ inflaton oscillations:
\begin{equation}
\frac{n_{\chi }}{n_{\varphi }}\sim \frac{k_{\ast }^{3}\exp \left( 2\pi \bar{%
\mu}_{k=0}N\right) }{m\Phi _{0}^{2}}\sim m^{1/2}\tilde{g}^{3/2}\cdot 3^{N},
\label{ir}
\end{equation}%
where $\Phi _{0}\sim O\left( 1\right) $ is the value of the inflaton
amplitude after the end of inflation. Since in the adiabatic regime the
effective mass of the $\chi -$particles is of order $\tilde{g}\Phi ,$ where $%
\Phi $ decreases in inverse proportion to $N,$ \ we also obtain an estimate
for the ratio of the energy densities:
\begin{equation}
\frac{\varepsilon _{\chi }}{\varepsilon _{\varphi }}\sim \frac{m_{\chi
}n_{\chi }}{mn_{\varphi }}\sim m^{-1/2}\tilde{g}^{5/2}N^{-1}3^{N}.
\label{ir1}
\end{equation}%
The formulae above fail when the energy density of the created particles
begins to exceed the energy density stored in the inflaton field. In fact,
at this time, the amplitude $\Phi \left( t\right) $ begins to decrease very
quickly because of the very efficient energy transfer from the inflaton to
the $\chi -$particles. Broad resonance is certainly over when $\Phi \left(
t\right) $ drops to the value $\Phi _{r}\sim m/\tilde{g},$ and we enter the
narrow resonance regime. For the coupling constant $m^{1/2}>\tilde{g}%
>O\left( 1\right) m,$ the number of the inflaton oscillation $N_{r}$ in the
broad resonance regime can be roughly estimated using the condition $%
\varepsilon _{\chi }\sim \varepsilon _{\varphi }:$
\begin{equation}
N_{r}\sim \left( 0.75\text{ to }2\right) \log _{3}m^{-1}.  \label{ios}
\end{equation}%
As an example, if $m\simeq 10^{13}$ GeV$,$ we have $N_{r}\simeq 10$ to $25$
for a wide range of the coupling constants $10^{-3}>\tilde{g}>10^{-6}$.
Taking into account that the total energy decays as $m^{2}\left( \Phi
_{0}/N\right) ^{2},$ we obtain
\begin{equation}
\frac{\varepsilon _{\varphi }}{\varepsilon _{\chi }+\varepsilon _{\varphi }}%
\sim \frac{m^{2}\Phi _{r}^{2}}{m^{2}\left( \Phi _{0}/N_{r}\right) ^{2}}\sim
N_{r}^{2}\left( \frac{m}{\tilde{g}\Phi _{0}}\right) ^{2},  \label{ienratio}
\end{equation}%
that is, the energy still stored in the inflaton field at the end of broad
resonance is only a small fraction of the total energy. In particular, for $%
m\simeq 10^{13}$ GeV$,$ this ratio varies in the range $10^{-6}$ to $O\left(
1\right) $ depending on the coupling constant $\tilde{g}.$

\begin{description}
\item \addtocounter{prob}{+1} \textbf{Problem \theprob.} Investigate
inflaton decay due to the three-leg interaction $g^{2}\varphi \chi ^{2}$ in
the strong coupling regime: $m>g>m^{2}/\Phi $.
\end{description}

\subsection{Implications}

It follows from the above considerations that broad parametric resonance can
play a very important role in the preheating phase. During only 15 to 25
oscillations of the inflaton, it can convert most of the inflaton energy
into other scalar particles. The most interesting aspect of this process is
that the effective mass and the momenta of the particles produced can exceed
the inflaton mass. For example, for $m\simeq 10^{14}$ GeV$,$ the effective
mass $m_{\chi }^{\text{eff}}=\tilde{g}\Phi \left\vert \cos \left( mt\right)
\right\vert $ can be as large as $10^{16}$ GeV$.$ Therefore, if the $\chi -$%
particles are coupled to bosonic and fermionic fields heavier than the
inflaton, then the inflaton may indirectly decay into these heavy particles.
This brings Grand Unification scales back into play. For instance, even if
inflation ends at low energy scales, preheating may rescue the GUT
baryogenesis models. Another potential outcome of the above mechanism is the
far-from-equilibrium production of topological defects after inflation.
Obviously their numbers must not conflict with observations and this leads
to cosmological bounds on admissible theories.

If, after the period of broad resonance, the slightest amount of the
inflaton remained $-$ given by (\ref{ienratio}) $-$ it would be a
cosmological disaster. Since the inflaton particles are nonrelativistic, if
they were present in any substantial amount, they would soon dominate and
leave us with a cold universe. Fortunately, these particles should easily
decay in the subsequent narrow resonance regime or as a result of elementary
particle decay. These decay channels thus become necessary ingredients of
the reheating theory.

The considerations of this section do not constitute a complete theory of
reheating. We have studied only elementary processes which could play a role
in producing a hot Friedmann universe. The final outcome of reheating must
be matter in thermal equilibrium. The particles which are produced in the
preheating processes are initially in a highly non-equilibrium state.
Numerical calculations show that as a result of their scatterings they
quickly reach local thermal equilibrium. Parametrizing the total preheating
and reheating time in terms of the inflaton oscillations number $N_{T},$ \
we obtain the following estimate for the reheating temperature:
\begin{equation}
T_{R}\sim \frac{m^{1/2}}{N_{T}^{1/2}N^{1/4}},  \label{irtemp}
\end{equation}%
where $N$ is the effective number degrees of freedom of the light fields at $%
T\sim T_{R}.$ Assuming that $N_{T}\sim 10^{6},$ and taking $N\sim 10^{2}$
and $m\simeq 10^{13}$ GeV$,$ we obtain $T_{R}\sim 10^{12}$ GeV$.$ This does
not mean, however, that we can ignore physics beyond this scale. As we have
already pointed out, non-equilibrium preheating processes can play a
nontrivial role.

Reheating is an important ingredient of inflationary cosmology. We have seen
that there is no general obstacle to arranging successful reheating. A
particle theory should be tested on its ability to realize reheating in
combination with baryogenesis. In this way, cosmology enables us to
preselect realistic particle physics theories beyond the Standard Model.

\section{\textquotedblleft Menu\textquotedblright\ of scenarios}

All we need for successful inflation is a scalar condensate
satisfying the slow-roll conditions. Building concrete scenarios
then becomes a \textquotedblleft technical\textquotedblright\
problem. Involving two or more scalar condensates, and assuming them
to be equally relevant during inflation, extends the number of
possibilities, but simultaneously diminishes the predictive power of
inflation. This especially concerns cosmological perturbations,
which are among the most important robust predictions of inflation.
Because inflation can be falsified experimentally (or more
accurately, observationally) only if it makes such predictions, we
consider only simple scenarios with a single inflaton component.
Fortunately all of them lead to very similar predictions which
differ only slightly in the details. This makes the significance of
a unique scenario, the one actually realized in nature, less
important. The situation here is very different from particle
physics, where the concrete models are as important as the ideas
behind them. This does not mean we do not need the correct scenario;
if one day it becomes available, we will be able to verify more
delicate predictions of inflation. However, even in the absence of
the true scenario, we can nonetheless verify observationally the
most important predictions of the stage of cosmic acceleration. The
purpose of this section is to give the reader a very brief guide to
the \textquotedblleft menu of scenarios\textquotedblright\ discussed
in the literature.

\textbf{Inflaton candidates}\textit{. }The first question which naturally
arises is \textquotedblleft what is the most realistic\ candidate for the
inflaton field?\textquotedblright . There are many because the only
requirement is that this candidate imitates a scalar condensate in the
slow-roll regime. This can be achieved by a fundamental scalar field or by a
fermionic condensate described in terms of an effective scalar field. This,
however, does not exhaust all possibilities. The scalar condensate can also
be imitated entirely within the theory of gravity itself. Einstein gravity
is only a low curvature limit of some more complicated theory whose action
contains higher powers of the curvature invariants, for example,
\begin{equation}
S=-%
\frac{1}{16\pi }\int \left( R+\alpha R^{2}+\beta R_{\mu \nu }R^{\mu \nu
}+\gamma R^{3}+...\right) \sqrt{-g}dx.  \label{iahd}
\end{equation}%
The quadratic and higher order terms can be either of fundamental origin or
they can arise as a result of vacuum polarization. The corresponding
dimensional coefficients in front of these terms are likely of Planckian
size. The theory with action (\ref{iahd}) can provide us with inflation.
This can easily be understood. Einstein gravity is the only metric theory in
four dimensions where the equations of motion are second order. Any
modification of the Einstein action introduces higher derivative terms. This
means that, in addition to the gravitational waves, the gravitational field
has extra degrees of freedom including, generically, a spin zero field.

\begin{description}
\item \addtocounter{prob}{+1} \textbf{Problem \theprob. }Consider a gravity
theory with metric $g_{\mu \nu }$ and action
\begin{equation}
S=\frac{1}{16\pi }\int f\left( R\right) \sqrt{-g}dx,  \label{iachd}
\end{equation}%
where $f\left( R\right) $ is an arbitrary function of the scalar curvature $%
R.$ Derive the following equations of motion:
\begin{equation}
\frac{\partial f}{\partial R}R_{\nu }^{\mu }-\frac{1}{2}\delta _{\nu }^{\mu
}f+\left( \frac{\partial f}{\partial R}\right) _{;\alpha }^{;\alpha }\delta
_{\nu }^{\mu }-\left( \frac{\partial f}{\partial R}\right) _{;\nu }^{;\mu
}=0.  \label{ife}
\end{equation}%
Verify that under the conformal transformation $g_{\mu \nu }\rightarrow
\tilde{g}_{\mu \nu }=Fg_{\mu \nu },$ the Ricci tensor and the scalar
curvature transform as
\begin{equation}
R_{\nu }^{\mu }\rightarrow \tilde{R}_{\nu }^{\mu }=F^{-1}R_{\nu }^{\mu
}-F^{-2}F_{;\nu }^{;\mu }-\frac{1}{2}F^{-2}F_{;\alpha }^{;\alpha }\delta
_{\nu }^{\mu }+\frac{3}{2}F^{-3}F_{;\nu }F^{;\mu },  \label{itrr}
\end{equation}%
\begin{equation}
R\rightarrow \tilde{R}=F^{-1}R-3F^{-2}F_{;\alpha }^{;\alpha }+\frac{3}{2}%
F^{-3}F_{;\alpha }F^{;\alpha }.  \label{itrs}
\end{equation}%
Introduce the \textquotedblleft scalar field\textquotedblright\
\begin{equation}
\varphi \equiv \sqrt{\frac{3}{16\pi }}\ln F\left( R\right) ,  \label{iscfhd}
\end{equation}%
and show that the equations
\begin{equation}
\tilde{R}_{\nu }^{\mu }-\frac{1}{2}\tilde{R}\delta _{\nu }^{\mu }=8\pi
\tilde{T}_{\nu }^{\mu }\left( \varphi \right) ,  \label{ieehd}
\end{equation}%
coincide with (\ref{ife}) if we set $F=\partial f/\partial R$ and take the
following potential for the scalar field:
\begin{equation}
V\left( \varphi \right) =\frac{1}{16\pi }\frac{f-R\partial f/\partial R}{%
\left( \partial f/\partial R\right) ^{2}}.  \label{ipothd}
\end{equation}

\item \addtocounter{prob}{+1} \textbf{Problem \theprob. }Study the
inflationary solutions in $R^{2}-$gravity:
\begin{equation}
S=-\frac{1}{16\pi }\int \left( R-\frac{1}{6M^{2}}R^{2}\right) \sqrt{-g}dx.
\label{ir2}
\end{equation}%
What is the physical meaning of the constant $M$?
\end{description}

Thus, the higher derivative gravity theory is conformally equivalent to
Einstein gravity with an extra scalar field. If the scalar field potential
satisfies the slow-roll conditions, then we have an inflationary solution in
the conformal frame for the metric $\tilde{g}_{\mu \nu }.$ One should not
confuse, however, the conformal metric with the original physical metric.
They generally describe manifolds with different geometries and the final
results must be interpreted in terms of the original metric. In our case the
use of the conformal transformation is a mathematical tool which simply
allows us to reduce the problem to one we have studied before. The conformal
metric is related to the physical metric $g_{\mu \nu }$ by a factor $F$,
which depends on the curvature invariants; it does not change significantly
during inflation. Therefore, we also have an inflationary solution in the
original physical frame.

So far we have been considering inflationary solutions due to the potential
of the scalar field. However, inflation can be realized even without a
potential term. It can occur in Born-Infeld type theories, where the action
depends nonlinearly on the kinetic energy of the scalar field. These
theories do not have higher derivative terms, but they have some other
peculiar properties.

\begin{description}
\item \addtocounter{prob}{+1} \textbf{Problem \theprob. }Consider a scalar
field with action
\begin{equation}
S=\int p\left( X,\varphi \right) \sqrt{-g}dx,  \label{igscf}
\end{equation}%
where $p$ is an arbitrary function of $\varphi $ and $X\equiv \dfrac{1}{2}%
\left( \partial _{\mu }\varphi \partial \varphi ^{\mu }\right) .$ Verify
that the energy-momentum tensor for this field can be written in the form
\begin{equation}
T_{\nu }^{\mu }=\left( \varepsilon +p\right) u^{\mu }u_{\nu }-p\delta _{\nu
}^{\mu },  \label{iemtn}
\end{equation}%
where the Lagrangian $p$ plays the role of the effective pressure and
\begin{equation}
\varepsilon =2X\frac{\partial p}{\partial X}-p,\text{ \ \ \ }u_{\nu }=\frac{%
\partial _{\nu }\varphi }{\sqrt{2X}}\text{\ }.\text{\ \ \ }  \label{ienvel}
\end{equation}%
If the Lagrangian $p$ satisfies the condition $X\partial p/\partial X\ll p$
for some range of $X$ and $\varphi ,$ then the equation of state is $%
p\approx -\varepsilon $ and we have an inflationary solution. Why is
inflation not satisfactory if $p$ depends only on $X$? Consider a general $%
p\left( X,\varphi \right) $ without an explicit potential term, that is, $%
p\rightarrow 0$ when $X\rightarrow 0.$ Formulate the conditions which this
function must satisfy to provide us with a slow-roll inflationary stage and
a graceful exit. The inflationary scenario based on the nontrivial
dependence of the Lagrangian on the kinetic term is called $k-$inflation.%
\index{inflation!$k$-inflation}%
\index{inflation!with kinetic term}
\end{description}

\textbf{Scenarios. }The simplest inflationary scenarios can be subdivided
into three classes. They correspond to the usual scalar field with a
potential, higher derivative gravity, and $k-$inflation. The cosmological
consequences of scenarios from the different classes are almost
indistinguishable $-$ they can exactly imitate each other. Within each
class, however, we can try to make further distinctions by addressing the
questions: a) what was before inflation? and b) how does a graceful exit to
a Friedmann stage occur? For our purpose it will be sufficient to consider
only the simplest case of a scalar field with canonical kinetic energy. The
potential can have different shapes, as shown in Fig.~7. The three cases
presented correspond to the so-called old, new and chaotic inflationary
scenarios. The first two names refer to their historical origins (see
\textquotedblleft Bibliography\textquotedblright ).

\textit{Old inflation} (see Fig.~7a) assumes that the scalar field arrives
at the local minimum of the potential at $\varphi =0$ as a result of a
supercooling of the initially hot universe. After that the universe
undergoes a stage of accelerated expansion with subsequent graceful exit via
bubble nucleation. It was clear from the very beginning that this scenario
could not provide a successful graceful exit because all the energy released
in a bubble is concentrated in its wall and the bubbles have no chance to
collide. This difficulty was avoided in the new inflationary scenario, a
scenario similar to a successful model in higher derivative gravity which
had previously been invented.

\begin{figure}
\begin{center}
\psfrag{V}[tr]{\small$V$}
\psfrag{f}[tr]{\small$\varphi$}
\psfrag{a}[t]{a)}
\psfrag{b}[t]{b)}
\psfrag{c}[t]{c)}
\includegraphics[width=0.9\textwidth,angle=0]{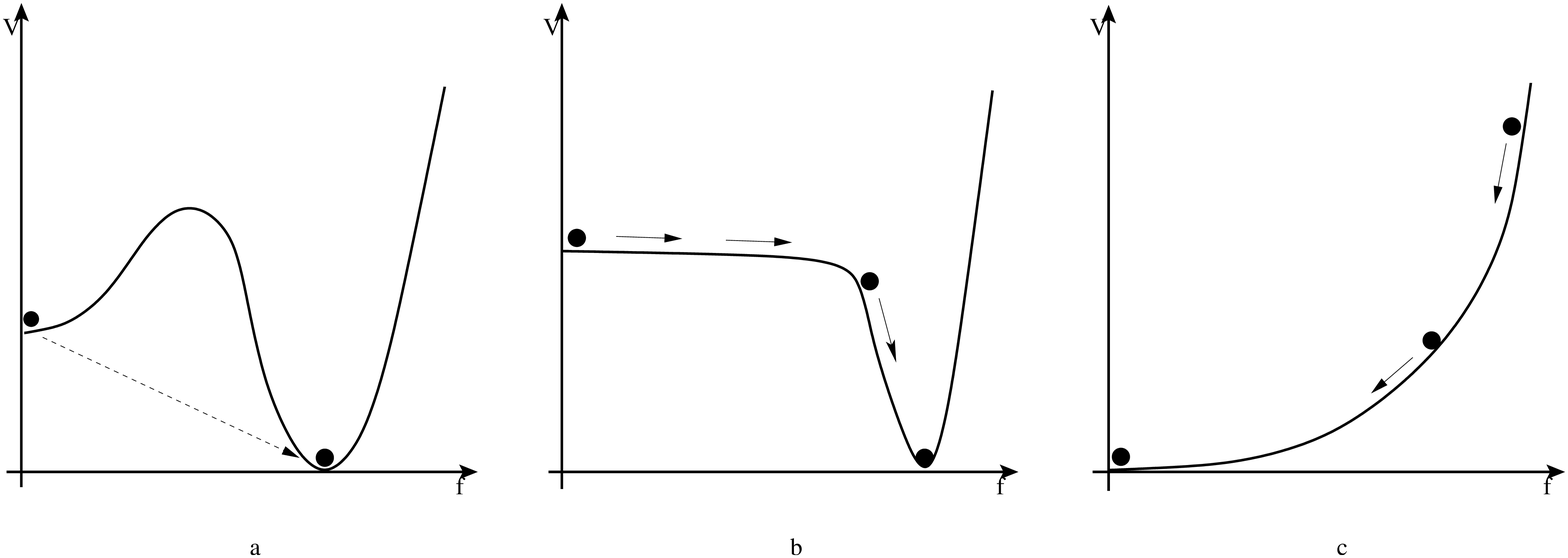}
\caption{~}
\end{center}
\end{figure}

\textit{New inflation} is based on a Coleman-Weinberg type potential (Fig.
7b). Because the potential is very flat and has a maximum at $\varphi =0,$
the scalar field escapes from the maximum not via tunneling, but due to the
quantum fluctuations. It then slowly rolls towards the global minimum where
the energy is released homogeneously in the whole space. Originally the
pre-inflationary state of the universe was taken to be thermal so that the
symmetry was restored due to thermal corrections. This was a justification
for the initial conditions of the scalar field. Later on it was realized
that the thermal initial state of the universe is quite unlikely, and so now
the original motivation for the initial conditions in the new inflationary
model seems to be false. Instead, the universe might be in a
\textquotedblleft self-reproducing\textquotedblright\ regime (for more
details see section 8.5).

\textit{Chaotic inflation} gives its name to the broadest possible class of
potentials satisfying the slow roll-conditions (Fig. 7c). We have considered
it in detail in the previous sections. The name \textit{chaotic} is related
to the possibility of having almost arbitrary initial conditions for the
scalar field. To be precise, this field must initially be larger than the
Planckian value but it is otherwise arbitrary. Indeed, it could have varied
from one spatial region to another and, as a result, the universe would have
a very complicated global structure. It could be very inhomogeneous on
scales much larger than the present horizon and extremely homogeneous on
\textquotedblleft small\textquotedblright\ scales corresponding to the
observable domain. We will see later that in the case of chaotic inflation,
quantum fluctuations lead to a self-reproducing universe.

Since chaotic inflation encompasses so many potentials, one might think it
worthwhile to consider special cases, for example, an exponential potential.
For an exponential potential, if the slow-roll conditions are satisfied
once, they are always satisfied. Therefore, it describes (power-law)
inflation without a graceful exit. To arrange a graceful exit we have to
\textquotedblleft damage\textquotedblright\ the potential. For two or more
scalar fields the number of options increases. Thus it is not helpful here
to go into the details of the different models.

In the absence of the underlying fundamental particle theory, one is free to
play with the potentials and invent more new scenarios. In this sense the
situation has changed since the time the importance of inflation was first
realized. In fact, in the 80s many people considered inflation a useful
application of the Grand Unified Theory that was believed to be known.
Besides solving\ the initial conditions problem, inflation also explained
why we do not have an overabundance of the monopoles that are an inevitable
consequence of a GUT. Either inflation ejects all previously created
monopoles, leaving less than one monopole per present horizon volume, or the
monopoles are never produced. The same argument applies to the heavy stable
particles that could be overproduced in the state of thermal equilibrium at
high temperatures. Many authors consider the solution of the monopole and
heavy particle problems to be as important as a solution of the initial
conditions problem. We would like to point out, however, that the initial
conditions problem is posed to us by nature, while the other problems are,
at present, not more than internal problems of theories beyond the Standard
Model. By solving these extra problems, inflation opens the door to theories
that would otherwise be prohibited by cosmology. Depending on one's
attitude, this is either a useful or damning achievement of inflation.

\textbf{De Sitter solution and inflation}. The last point we would like to
make concerns the role of a cosmological constant and a pure de Sitter
solution for inflation. We have already said that the pure de Sitter
solution cannot provide us with a model with a graceful exit. Even the
notion of expansion is not unambiguously defined in de Sitter space. We saw
in section 1.3.6 that this space has the same symmetry group as Minkowski
space. It is spatially homogeneous and time translation invariant. Therefore
any space-like surface is a hypersurface of constant energy. To characterize
an expansion we can use not only $k=0,\pm 1$ Friedmann coordinates but also,
for example, \textquotedblleft static coordinates\textquotedblright\ (see
Problem 2.7), which describe an expanding space outside the event horizon.
In all these cases the three-geometries of constant time hypersurfaces are
very different. These differences, however, simply characterize the
different slicings of the perfectly symmetrical space and there is no
obvious preferable choice for the coordinates.

\begin{figure}
\begin{center}
\psfrag{H}[]{\small our history}
\psfrag{h1}[l]{\small$\bar{\eta}=const \ (k=0)$}
\psfrag{h2}[l]{\small$\eta=const \ (k=1)$}
\psfrag{x}[t]{\small$$}
\psfrag{h}[r]{\small$$}

\includegraphics[width=3in,angle=0]{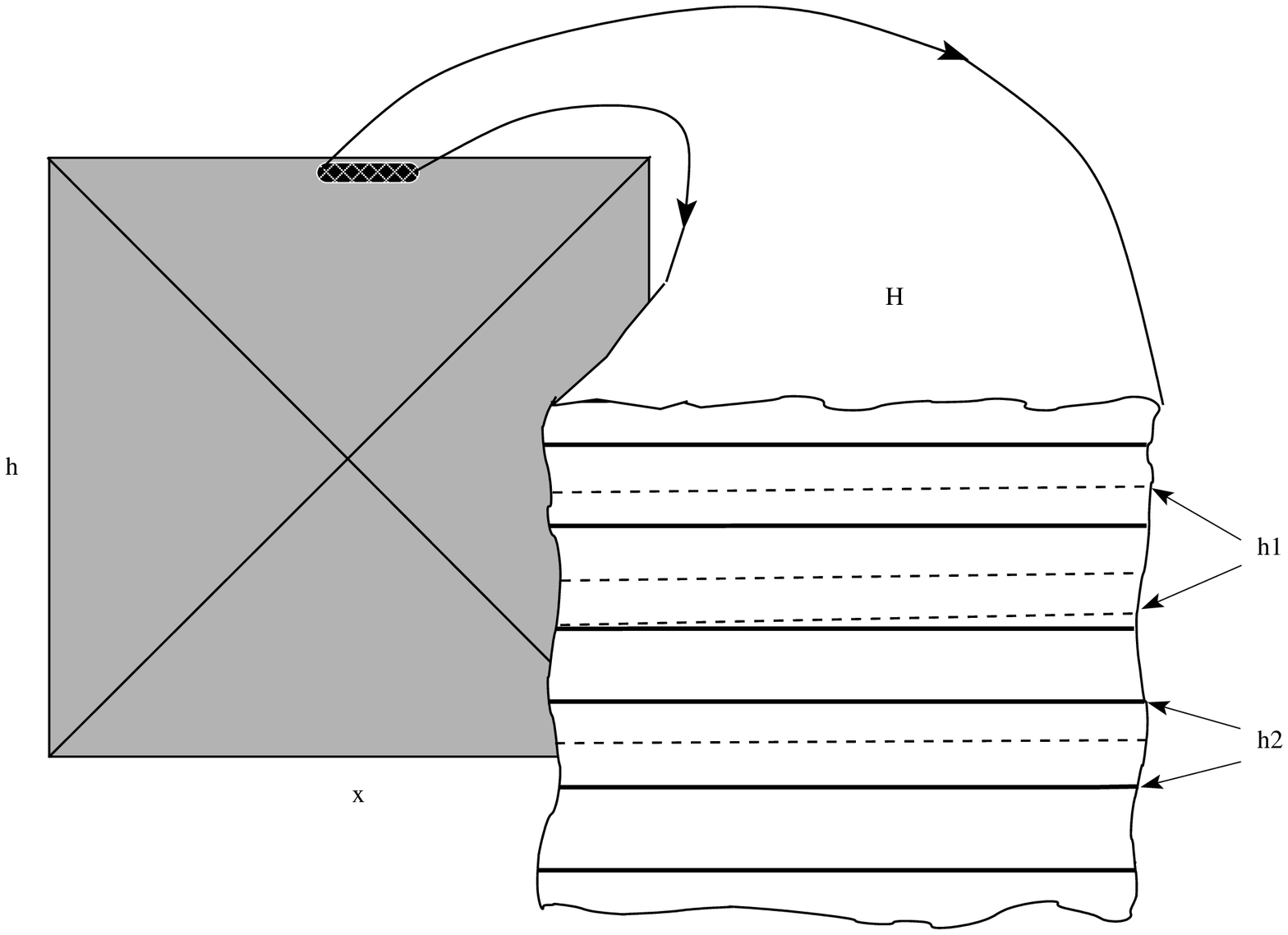}
\caption{~}
\end{center}
\end{figure}

It is important therefore that inflation is never realized by a pure de
Sitter solution. There must be deviations from the vacuum equation of state,
which finally determine the \textquotedblleft
hypersurface\textquotedblright\ of transition to the hot universe. The de
Sitter universe is still, however, a very useful zeroth order approximation
for nearly all inflationary models. In fact, the effective equation of state
must satisfy the condition $\varepsilon +3p<0$ for at least 75 e-folds. This
is generally possible only if during most of the time we have $p\approx
-\varepsilon $ to a rather high accuracy. Therefore, one can use the
language of constant time hypersurfaces defined in various coordinate
systems in de Sitter space. Our earlier considerations show that the
transition from inflation to the Friedmann universe occurs along a
hypersurface of constant time in the expanding isotropic coordinates $\left(
\eta =const\right) $, but not along $r=const$ hypersurface in the
\textquotedblleft static coordinates.\textquotedblright\ The next question
is, out of the three possible isotropic coordinate systems ($k=0,\pm 1$),
which must be used to match the de Sitter space to the Friedmann universe?
Depending on the answer to this question, we obtain flat, open or closed
Friedmann universes. It turns out, however, that this answer seems not to be
relevant for the observable domain of the universe. In fact, if inflation
lasts more than 75 e-folds, the observable part of the universe corresponds
only to a tiny piece of the matched global conformal diagrams for de Sitter
and Friedmann universes. This piece is located near the upper border of the
conformal diagram for de Sitter space and the lower border for the flat,
open or closed Friedmann universes (Fig.~8), where the difference between
the hypersurfaces of constant time for flat, open or closed cases is
negligibly small. After a graceful exit we obtain a very large domain of the
Friedmann universe with incredibly small flatness and this domain covers all
present observable scales. The global structure of the universe on scales
much larger than the present horizon is not relevant for an observer $-$ at
least not for the next hundred billion years. In the next section we will
see that the issue of the global structure is complicated by quantum
fluctuations. These fluctuations are amplified during inflation and as a
result the hypersurface of transition has \textquotedblleft
wrinkles.\textquotedblright\ The wrinkles are rather small on scales
corresponding to the observable universe but they become huge on the very
large scales. Hence, globally the universe is very different from the
Friedmann space and the question about the spatial curvature of the whole
universe no longer makes sense. It also follows that the global properties
of an exact de Sitter solution have no relevance for the real physical
universe.

\end{document}